\documentclass[preprint,12pt]{elsarticle}

\usepackage{graphicx}
\usepackage[pdftex,dvipsnames,usenames]{color}		
\usepackage{subfig}

\usepackage{amssymb} 	

 \usepackage{lineno}
 \usepackage{alphalph}

\usepackage{soul}   

\def\VYP#1#2#3{{\bf #1}, #3 (#2)}  

\def\PLB#1#2#3{{\it Phys.~Lett.~B}~\VYP{#1}{#2}{#3}}

\def\PRD#1#2#3{{\it Phys.~Rev.~D}~\VYP{#1}{#2}{#3}}
\def\PRL#1#2#3{{\it Phys.~Rev.~Lett.}~\VYP{#1}{#2}{#3}}

\def\NIM#1#2#3{{\it Nucl.~Inst.~and~Meth.~A}~\VYP{#1}{#2}{#3}}
\newcommand{\etal}{\mbox{\textit et al.}}                       %

\def\beq{\begin{equation}}
\def\eeq{\end{equation}}
\def\bea{\begin{eqnarray}}
\def\eea{\end{eqnarray}}
\def\eqref#1{Eq.~(\ref{eq:#1})}
\def\eqlab#1{\label{eq:#1}}
\newcommand*{\tabref}[1]{Table~\ref{tbl:#1}}
\newcommand*{\tablab}[1]{\label{tbl:#1}}
\newcommand*{\figref}[1]{Fig.~\ref{fig:#1}}
\newcommand*{\figlab}[1]{\label{fig:#1}}

\newcommand*{\secref}[1]{Sec.~\ref{sec:#1}}
\newcommand*{\seclab}[1]{\label{sec:#1}}

\newcommand{\Omit}[1]{}

\begin{document}

\begin{frontmatter}


\title{Optimized Trigger for Ultra-High-Energy Cosmic-Ray and Neutrino Observations with the Low Frequency Radio Array}



\author[KVI,VUB,Can]{K. Singh}
 \address[KVI]{Kernfysisch Versneller Instituut, University of Groningen, 9747 AA Groningen, The Netherlands}
 \address[VUB]{Vrije Universiteit Brussel, Dienst ELEM, B-1050 Brussels, Belgium}
 \address[Can]{Department of Physics, University of Alberta, Edmonton, AB, T6G 2G7, Canada\fnref{label2}}
 \author[KVI]{M.~Mevius} 
 \author[KVI]{O.~Scholten} 

\author[mpifr]{J.M.~Anderson}
 \address[mpifr]{Max-Planck-Institut f\"{u}r Radioastronomie, Auf dem H\"{u}gel 69, 53121 Bonn, Germany}
\author[Astr]{A.~van~Ardenne}
 \address[Astr]{ASTRON, Oude Hoogeveensedijk 4, 7991 PD Dwingeloo, The Netherlands}
\author[Astr]{M.~Arts}
\author[Astr,kapteyn]{M.~Avruch}
\author[Astr]{A.~Asgekar}
\author[soton]{M.~Bell}
 \address[soton]{School of Physics and Astronomy, University of Southampton, Southampton, SO17 1BJ, UK}
\author[Astr]{P.~Bennema}
\author[Astr]{M.~Bentum}
\author[cfa,kapteyn]{G.~Bernadi}
 \address[cfa]{Center for Astrophysics, Harvard University, USA}
 \address[kapteyn]{Kapteyn Astronomical Institute, PO Box 800, 9700 AV Groningen, The Netherlands}
% ...


\begin{abstract}
When an ultra-high energy neutrino or cosmic ray strikes the Lunar surface a radio-frequency pulse is emitted. We plan to use the LOFAR radio telescope to detect these pulses. In this work we propose an efficient trigger implementation for LOFAR  optimized for the observation of short radio pulses.
\end{abstract}

\tableofcontents


\end{frontmatter}

 \linenumbers

\section{Introduction}

Ultra-High Energy (UHE) cosmic-ray particles are a source of much speculation.  Particles with more than $10^{20}$\,eV of energy have been observed, but the source of these particles is an open question in astroparticle physics.   Such energetic particles are extremely rare; their flux on Earth is less than $ 1 $\,km$^{-2}$century$^{-1}$.  This low flux makes it difficult to determine the origin of these particles.  They may be accelerated by shock waves in Active Galactic Nuclei (AGN)~\cite{Protheroe04}, but it is also possible that they are created by annihilating or decaying dark-matter particles~\cite{Berezinsky10}. We do know that these UHE cosmic rays will not be bent appreciably by the galactic magnetic field, because their high momentum gives them high magnetic rigidity.  Furthermore, due to the GZK effect~\cite{GZK}, the sources of the UHE cosmic rays we do detect have to be close to Earth, a distance of the order of 50\,MPc or less, as it prevents us from detecting UHE cosmic rays from distant sources.

There is an alternative approach to finding the sources of UHE cosmic rays.  Instead of detecting the cosmic rays directly, we aim to detect the neutrinos that are produced at their creation sites or in transport through their interaction with the cosmic microwave background~\cite{GZK}, known as the GZK effect.  These neutrinos will carry most of the energy of the original cosmic ray, but are almost unaffected by the intergalactic medium, and thus carry direct information about the UHE cosmic rays from distant sources.

Because of their limited interactions, neutrinos are very difficult to detect.  To measure the small flux of UHE neutrinos, it is necessary to use detectors with an extremely large acceptance. Such detectors include the Pierre Auger Observatory~\cite{PAO}, ICECUBE~\cite{icecube}, ANITA~\cite{anita10}, FORTE~\cite{forte} and KM3Net~\cite{KM3NET}.

Celestial bodies can serve as large-acceptance detectors.  In 1989, Dagkesamanskii and Zhelenznykh~\cite{Dag89} proposed using the Askaryan effect~\cite{Askar} to measure the flux of UHE neutrinos impinging on the Moon.  The Moon offers an acceptance area on the order of $10^7$\,km$^2$, far larger than any man-made structure. Having such a large acceptance allows for sensitive measurements of the flux of these UHE neutrinos and cosmic rays.  Based on Dagesamanskii and Zhelenznykh's concept, experiments have been carried out at the Parkes~\cite{Han96,James07},  Goldstone~\cite{Gor04}, Kalyazin~\cite{Ber05},  and recently at the VLA~\cite{Jaeger10} telescopes.  These experiments have looked for short radio pulses near the frequency where the intensity of the Askaryan effect is expected to reach its maximum.  It may be advantageous to look for pulses at lower frequencies, where the angular spread of the emission around the Cherenkov angle is larger.  This results in an increase in detection sensitivity~\cite{Sch06} for three reasons: for a much larger range of incident angles the radio waves will reach Earth, internal reflection at the Lunar surface is of lesser importance, and the absorption length increases, which means that the waves emitted by neutrino-induced showers at greater depth will still be detectable.
It was shown~\cite{Sch06} that the optimum frequency-window for this observation is around 100--200 MHz.
To perform observations of narrow transients in this frequency band a new program was initiated called NuMoon. Initially the Westerbork Synthesis Radio Telescope (WSRT) has been used to make such observations at frequencies near 150 MHz.
These observations have been used to improve the flux-limit for UHE neutrinos~\cite{Sch09NM,Bui10} by about an order of magnitude. We aim to further improve this result by using LOFAR (LOw Frequency ARray)~\cite{Hoer09}.  With LOFAR's larger collecting area and wider frequency range, a 25 times higher sensitivity for the detection of UHE particles is within reach~\cite{Singh08}.

The main issue for the NuMoon observations with LOFAR is dealing with the high data rate.
The data rate of the raw time-series is about 1~TB/s. Even if only 1~ms of data is stored per event, this still creates a high load on the data transmission lines and storage devices at the CEntral Processor facility (CEP) and necessitates the implementation of a very efficient trigger algorithm. It is crucial to reduce the number of false detection events, since a single event consists of about 1.6 GB of data.  The triggering criteria must be optimized so that false detection events occur infrequently, but real events are not missed. It should be realized that this last condition is essential since only the triggered data are stored and available for later processing. If the trigger condition is too constraining,  we would not be sensitive to pulses that could easily be distinguished from a noise signal in an offline analysis using the full capability of LOFAR.
The construction of the trigger algorithm is the subject of this work.

The remote stations and the international stations of LOFAR (see \secref{LOFAR}) are important to offline analysis of the detected events.  There are two chief benefits to using remote and international stations.  One, because of the increased collecting area, the signal-to-background ratio will be improved when these stations are used to form tied-array beams in an offline analysis. 
Two, the pointing resolution of LOFAR is much better when well-separated stations are contributing data because of the large interferometric baseline of these stations.  Improved pointing resolution increases the efficiency of the anti-coincidence criterion.  It also gives better information about the origins of genuine pulses. Knowing the place on the Moon where the signal originates from allows for an accurate accounting of the Lunar terrain in  simulations of the signal.

The general structure of this paper is as follows, we start by presenting a general outline of LOFAR in \secref{LOFAR} with emphasis on the aspects which are relevant for the construction of the trigger.  For technical reasons, for the construction of an optimized trigger algorithm, only part of the full band width may be used. In \secref{trigger} we discuss the different alternatives for selecting the part of the band that will be used. The pulse-search algorithm is presented in \secref{PulseSearch} in conjunction with the procedure to optimize it. The signals arriving at the Earth from the Moon pass through the ionosphere, which induces a dispersion that can be corrected for to a large extent as discussed in \secref{IonDis}. The complete simulation, including the effects of a distributed antenna system, is presented in \secref{Beam}.
In \secref{limits} the attainable flux limits are given for UHE neutrinos and cosmic rays, given the sensitivity of the trigger algorithm.

\section{Trigger implementation at LOFAR}\seclab{LOFAR}

LOFAR is a multi-purpose sensor array~\cite{lofar,Falcke} whose main application is radio astronomy.  As the name suggests, LOFAR is sensitive to low frequencies (
10--240\,MHz).  It is a distributed radio-interferometric array consisting of many low-cost antennas. 
These antennas are organized into many separate array stations, and 40 of these stations are located in the northeastern Netherlands. About half (24 when the array is completed) of these Dutch stations form the ``core" of LOFAR, and these core stations are clustered into an area 2\,km in diameter.  The other 16 Dutch stations are called remote stations, and they are located within 80 km of the core. Additionally, international stations have been constructed or are planned in various other European countries.
These countries include Germany, the UK, France and Sweden. 
The use of the international stations gives LOFAR an interferometric baseline of approximately 1500\,km.  The maximal interferometric baseline within the Netherlands is on the order of 100\,km.

LOFAR works with two distinct antenna types, Low Band Antennas (LBA), which operate between 10 and 80 MHz, and High Band Antennas (HBA), which operate between 110 and 240 MHz.  In the present investigation, we are interested in the  in the 110--190 MHz region of the HBA antennas.
These are bow-tie-shaped dual-dipole antennas, which are assembled in a 4X4 grid measuring $5$\,m$\times 5$\,m.
For each core station of LOFAR, the HBA antennas are grouped into two sub-fields, each with 24 HBA-tiles and a diameter of 35\,m.  The distance between the two groups is about 129\,m.  A remote station consists of a single group of 48 HBA-tiles.  This group has a total diameter of about 50m.  An international station consists of a single group of 96 HBA-tiles with a total diameter of about 62\,m.

The signals received by all antennas of a single HBA tile are added by an analogue beamformer. Subsequently, the signals of all tiles of a single station are collected, and appropriate phase-delays are applied to form the station beams.  These digitally synthesized station beams are equivalent to the beam of a single dish of a traditional radio telescope.  Each LOFAR station has a 10 Gbit/s connection to CEP with a real data rate between 3.2\,Gbit/s per station.
The CEP is an IBM Blue Gene/P supercomputer and additional off-line clusters, located in Groningen, and is responsible for collecting and processing the data from the LOFAR stations.  The beams of the core stations are added in phase to form tied-array beams online at CEP~\cite{Stappers2011, Mol2011}.  The use of tied-array beams improves the pointing resolution of LOFAR, since the core stations have an interferometric baseline on the order of 2\,km. In addition, tied array beam forming, by summing station beams in phase, increases the effective area and thus the signal to noise ratio.

In parallel to this online data-processing, the digitized raw data of each tile 
are stored in  ring buffers, the Transient Buffer Boards (TBBs), at the station.
The raw data stored in the TBBs can be accessed for offline processing~\cite{Singh08}.  Each TBB stores the data from 8 dual polarized tiles for 1.3\,s where there are advanced plans to extend this to 5.2\,s.
The boards will upload these data to CEP when triggered by the pulse detection software.

The observation mode of LOFAR to detect cosmic rays and neutrinos at energies above $10^{21}$\,eV through their impacts on the Lunar surface is called the Ultra-High-Energy Particle (UHEP)-mode or the NuMoon-mode. In this mode digital beams, pointing to different spots on the Lunar surface, will be formed using all HBA fields 
 ($24\times 2\times 24$ tiles) of the core stations of LOFAR.
These data will be searched for short pulses. 
When a pulse is found, a trigger is sent to the TBBs. The raw data in the TBBs are then sent to CEP for storage and later offline processing.  This offline processing increases NuMoon's sensitivity and reduces the occurrence of false detection events.

\subsection{Data flow}

\begin{figure*}[!ht]
  \centering
  \includegraphics[width=.99\linewidth]{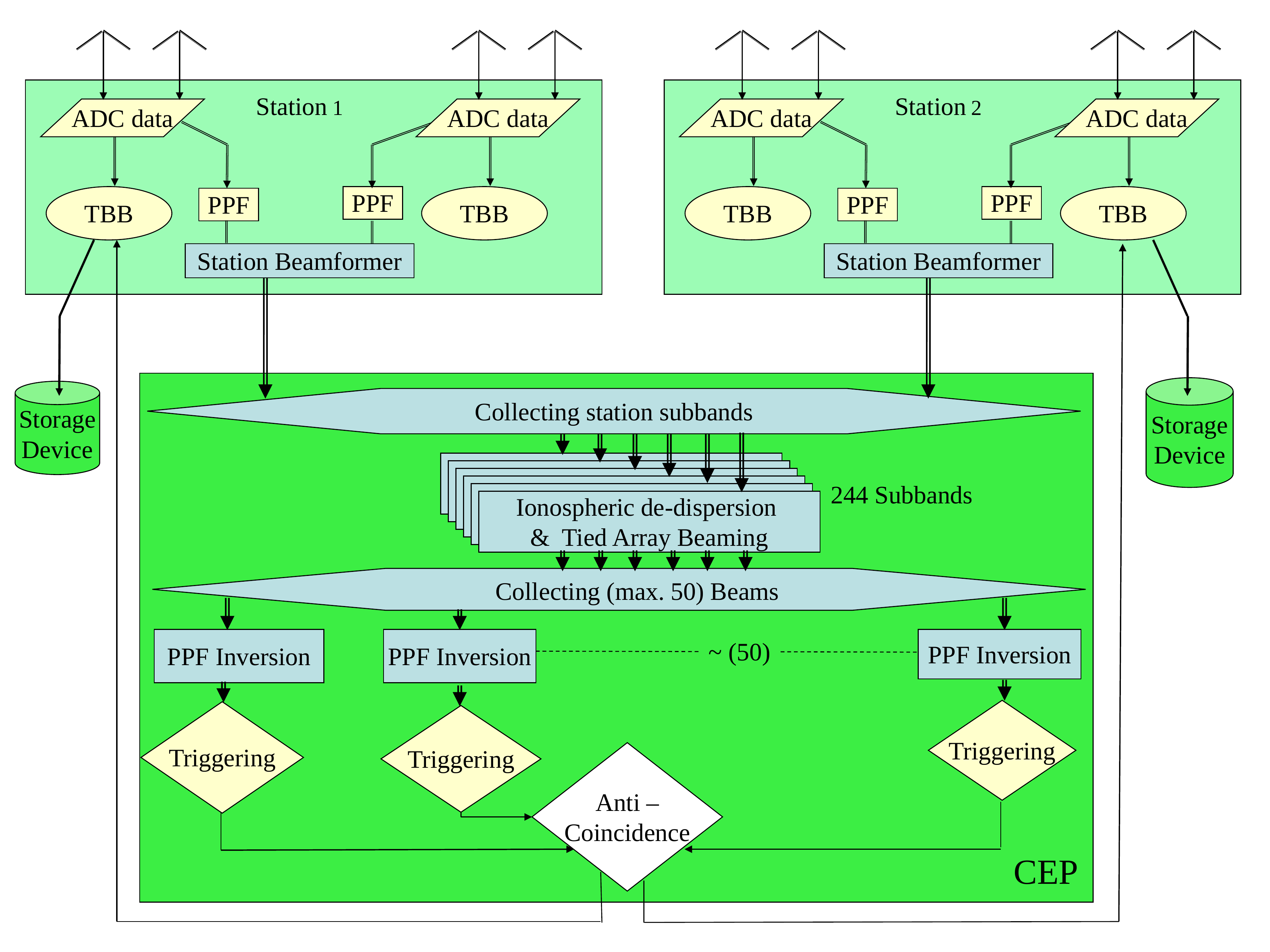}
 \caption{Online signal processing of LOFAR in the NuMoon pipeline~\cite{Singh09}.}
  \figlab{flowchart}
 \end{figure*}

The data flow through the system starting at the antennas is schematically depicted in \figref{flowchart}. The main structures indicated are the many stations in the field, schematically shown by the two boxes in the upper half of the figure. Each station receives the signals from the HBA-tiles of the station where each HBA-tile contains 16 dual antennas. In the station electronics the analog signals of each tile are sampled at 200\,MHz and converted to 12 bit digital samples. The digitized data are stored on a ring buffer for possible later processing. In addition the digitized signals are fed into a PolyPhase Filter (PPF) that also performs a Fast Fourier Transform (FFT) resulting in 512 frequency channels (subbands). The merits of the PPF are discussed in detail in the appendix. In the station beamformer the subbands of all tiles of a single HBA field are added in phase to form a single station beam.
The phase-masks necessary for forming the station beams are recalculated by local control units every second for the source (the Moon) under observation.
Each station beam is sent to CEP in the form of 244 frequency channels (subbands) as indicated by the heavy black arrows connecting the stations and CEP, corresponding to approximately half the available bandwidth.

At CEP the data of all stations are collected and a correction is applied to compensate for the ionospheric dispersion of the signal.
The massive parallel processing capability of CEP is used to apply  station-dependent phase shifts to form 50 tied-array beams for each of the 244 subbands.  Simulations show that 50 tied-array beams are sufficient to cover the full Lunar surface as discussed in \secref{Beam}.  These 50 beams are aimed at different patches of the visible Lunar surface.
All the subbands of a single beam at a single computing node are then collected and the data are transformed back into the time-domain.  In this step, the effect of the PPF is inverted (PPF inversion, see Appendix A).

Once the data have been converted back to the time-domain, each beam is searched for suitable pulses. The design of an efficient search procedure is the main subject of this work. In practice, many pulses will be due to transient noise.  It is necessary to have an efficient procedure to distinguish the noise pulses from genuine cosmic ray events (hereafter ``genuine" pulses will refer to events caused by cosmic ray and neutrino impacts on the Moon).  We can make use of the fact that genuine pulses come from a very localized spot on the Lunar surface.
A genuine pulse will thus be detected in one or at most a few adjacent beams.
In similar observations using the Westerbork Synthesis Radio Telescope~\cite{Sch09NM,Bui10}, as well as the Parkes telescope~\cite{Bra10}, it has been found that putting an anti-coincidence requirement between the beams is an efficient means of suppressing transient-noise triggers.  The NuMoon pipeline at LOFAR will incorporate such an anti-coincidence requirement in its triggering criteria. In implementing this trigger, care must be taken with the side-lobe sensitivities of the beams which are investigated in \secref{Beam}. Recall that triggering causes the TBBs to upload large amounts of data to CEP and results in system dead time.

In our analysis, we have simulated each block in the data-processing chain of \figref{flowchart}.  In this way, we estimate the total pulse-detection efficiency for the NuMoon observing mode of LOFAR.

\section{Filtering}\seclab{trigger}

Because of limitations in communication bandwidth and processing power at the station level, only 244 of the 512 subbands can be processed online. The data from these subbands will be sent to CEP, and CEP will search these data for signs of a NuMoon pulse. If a pulse is found, CEP will trigger a data-upload.  To reduce the occurrence of false triggers, we must select 244 subbands that are free from Narrowband Radio-Frequency Interference (NRFI). In \secref{RFIMitigation} it is shown that this can easily be done by introducing an NRFI mask. It will however be necessary to monitor the NRFI situation so that the NRFI mask can be adjusted if new NRFI lines appear. Broadband Radio-Frequency Interference, also called transient noise in this work, is much harder to eliminate as it may resemble the short pulses we search for.  This transient noise is addressed in \secref{RealData}.

Once NRFI lines have been excluded, we have some freedom to make a selection of the remaining subbands. Two considerations enter here. One is that the sensitivity for Lunar pulses is highest at the lowest frequencies, as is re-iterated in \secref{OW}. A second criterion is that the subband selection will affect the structure of the time-domain data that is reconstructed at CEP. In turn, this structure affects how well CEP detects NuMoon pulses.

\subsection {Narrowband radio frequency interference mitigation}\seclab{RFIMitigation}

In \figref{rfimitigation} a typical frequency spectrum of a single HBA tile of LOFAR is shown as was recently measured. Apart from the strong, narrow radio-frequency line at $169.65$\,MHz~\cite{C2000}, there are a few other narrow lines in the frequency spectrum.  These other lines are not always seen in the spectrum.  Nonetheless, they must be filtered out of the NuMoon data, since they contain an appreciable fraction of the power in the bandwidth when they are present.  This filtering is referred to as  NRFI mitigation, where NRFI stands for Radio Frequency Interference at a well defined frequency. 

\begin{figure}[!ht]
 \centering
 \includegraphics[width=.5\textwidth]{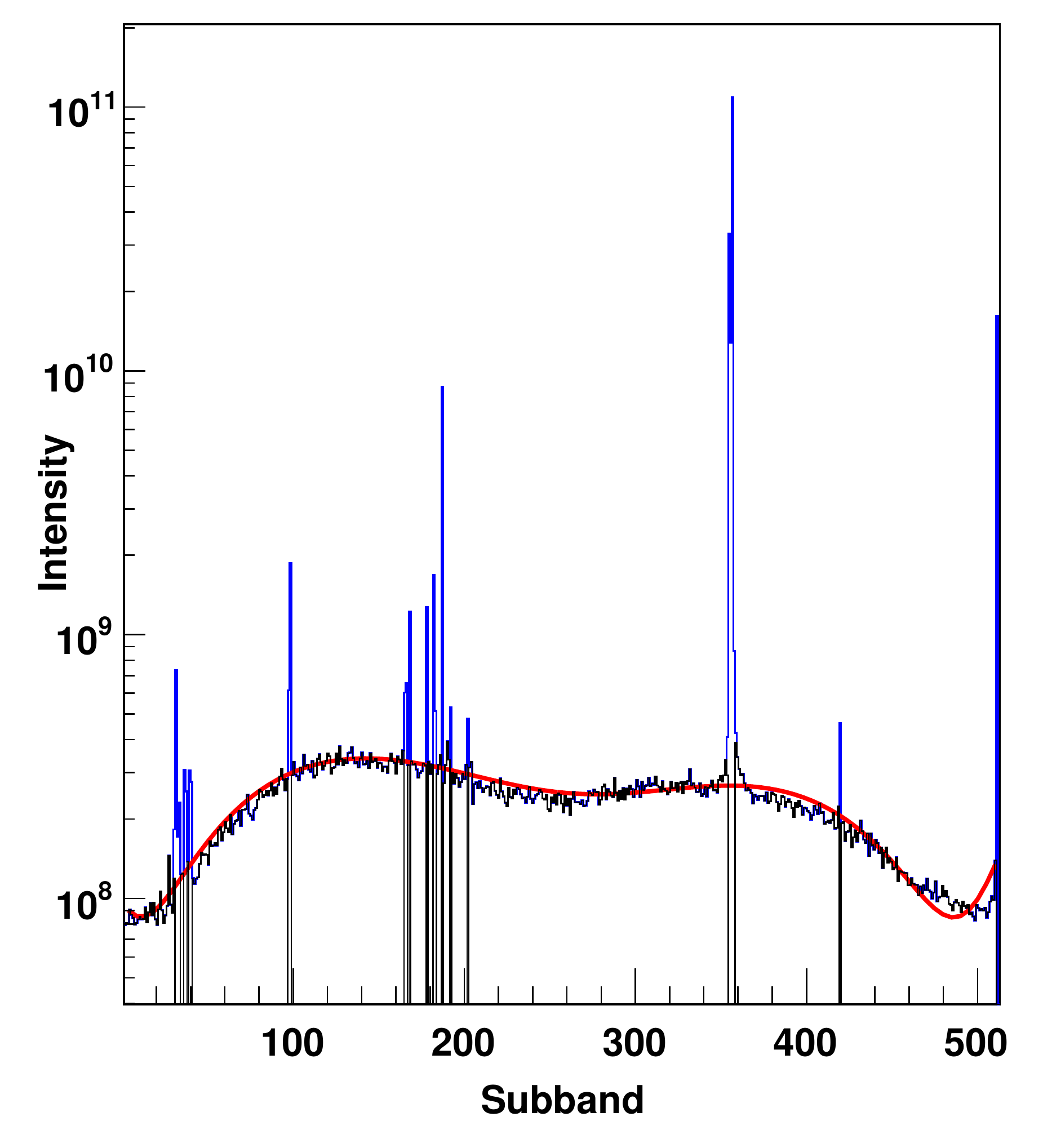}
 \caption{The blue curve shows the intensity (in arbitrary units) per subband, summed over 1\,ms for a single HBA tile. The fitted polynomial is shown in red. The NRFI subtracted spectrum is shown in
   black. Each subband has a width of 
 195.3125 kHz and the first (last) correspond to $100$ MHz ($200$ MHz). The $354^{th}$ subband contains the strong $169.65$\,MHz signal.}
 \figlab{rfimitigation}
\end{figure}

One possible NRFI mitigation procedure is as follows. First, the frequency spectrum for one polarization (blue curve in \figref{rfimitigation}) is summed over one block of data (1\,ms) consisting of 200 pages, where each page of 5\,$\mu$s contains 1024 time samples. This summed spectrum is fitted with a $6^{th}$ order polynomial (red curve in \figref{rfimitigation}).
The frequency subbands containing a power exceeding the fit by more than $50\%$ are marked as NRFI lines. For this reason the fit does not have to be very detailed. The subbands near the edges of the bandwidth are suppressed by the filters and are for this reason excluded from the analysis. The contents of these subbands are set to zero, giving the black curve in \figref{rfimitigation}. This type of NRFI mitigation is known as masking.  NRFI lines are not constant, so the mask must be updated once every few seconds or so. This procedure was applied in the analysis presented in Ref.~\cite{Sch09NM,Bui10}.

\begin{figure}[!ht]
 \centering
 \includegraphics[width=.5\textwidth]{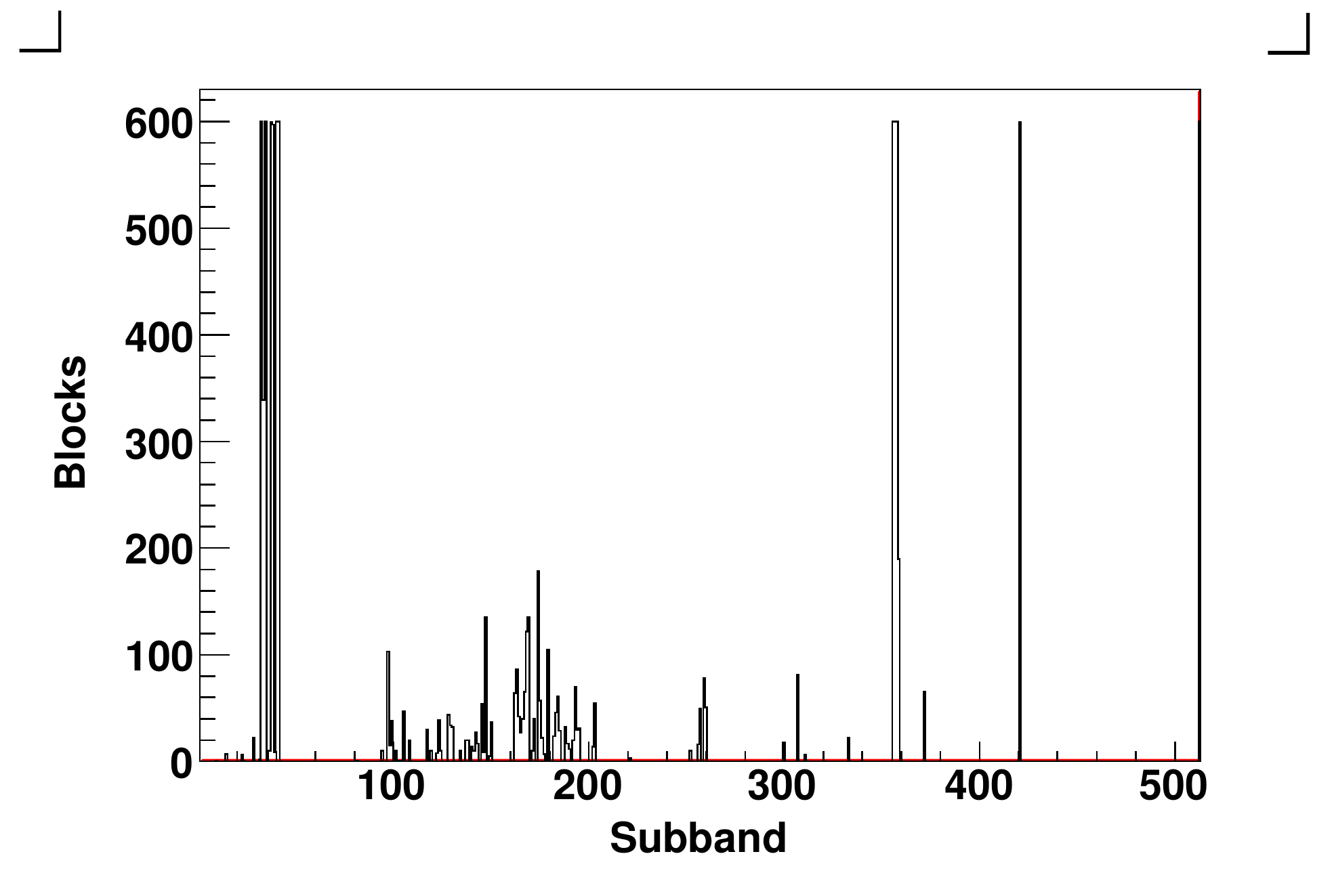}
 \caption{The number of blocks of 1 ms in which a subband is corrupted by NRFI is plotted vs.\ subband number for 0.6~seconds of single HBA tile data.}
 \figlab{rfi_situation}
\end{figure}

For LOFAR, the NRFI mitigation needs to be done online.  We thus have to minimize the extra latency in the data processing on CEP, which implies that online we have to work with a pre-defined frequency mask and cannot use the procedure outlined above.  For this reason we have performed an offline check on the HBA data, using a mask that updates regularly.
\figref{rfi_situation} shows the number of blocks of data (of 1\,ms each) in which a subband is dominated by NRFI. This is done for all $512$ frequency bins (subbands) for 0.6\,second of data (600 blocks) of a single tile as obtained from the TBB. The $169.65$\,MHz signal appears in every page and the count for this line reaches the maximum of $600$ in the $354^{th}$ subband. Also the $420^{th}$ subband is strongly affected by NRFI. In the first few subbands there are also continual NRFI lines, as shown in \figref{rfi_situation}, however, the filter gain is low for these subbands (see \figref{rfimitigation}).
Please note that the spectrum shown in \figref{rfimitigation} is only one of many needed to generate \figref{rfi_situation}.
There is also a region between the $100^{th}$ and $200^{th}$ subband where at times NRFI lines appear.
It should be noted that we have checked the NRFI situation at different hours of the day (at 5\,AM, 11\,AM, 5\,PM and 11\,PM) and the worst situation, occurring at 11 in the morning, is given in \figref{rfi_situation}.

\Omit{
\begin{table}[!ht]
 \caption{NRFI frequencies in MHz used in the NRFI mask.}
 \tablab{RFI-table}
 \begin{center}
 \begin{tabular}{|c|c|c|c|c|}
 \hline
 105.96 & 106.15 & 106.35 & 106.54 & 106.74 \\
 106.93 & 107.13 & 107.32 & 107.52 & 107.71  \\
 107.91 & 131.74 & 133.89 & 169.24 & 169.63 \\
 169.82 & 170.02 & 181.93  &   &   \\
\hline
 \end{tabular}
 \end{center}
 \end{table}
as of 3/19/11: 0    30    31    32    33    34    35    36    37    38    39    40   162   173   354   356   357   358   419   511}
Based on the observations presented in \figref{rfimitigation} and \figref{rfi_situation} we have made an NRFI-mask excluding subband with central frequencies at 131.64, 133.79, 169.14, 169.53, 169.72, 196.92 and 181.83\,MHz. 
It should be noted that, because the NRFI frequencies depend on the time of the day, the NRFI mask needs regular updating in the actual observations. The relation between subband number $n$ and its central frequency is given by $\nu=100$\,MHz$\;+\;n\,d\nu$ with $d\nu=195.3125$\,kHz. 
In addition we have also excluded the low-gain bands from our analysis with frequencies below 110\,MHz and above 190\,MHz.

\subsection{Optimum window}\seclab{OW}

In the design of the optimum trigger condition two aspects need to be considered.
The first is the spreading of the pulse in the time domain due to the partial bandwidth and due to ionospheric dispersion. The second important aspect is the variation of the sensitivity of LOFAR over the frequency regime.

The effective area of the HBA tiles of LOFAR can be written as~\cite{LofarAntenna,LofarAntenna2}
\beq
 A_\mathrm{eff}=\min(\lambda^2/3,1.5625)\,\mathrm{m}^2 \;,
\eeq
where the change over from a constant to a frequency dependent effective area occurs at a frequency of 138 MHz. The other important ingredient is the system temperature $T_\mathrm{sys}=T_\mathrm{sky} + T_\mathrm{inst}$, where $T_\mathrm{inst}\approx 200$\,K is the instrumental temperature, and the sky temperature can be written as
\beq
T_\mathrm{sky}=T_{s0} \left ( {\lambda}\over{1~\mathrm{m}} \right )^{2.55} \;, 
\eeq
where $T_{s0}=60 \pm 20$\,K and $\lambda$ has units of m. With these two ingredients the system equivalent flux density (SEFD) for Nyquist sampling can be expressed as
\beq
S_\mathrm{sys}= { 2 \eta k T_{\mathrm{sys}} \over A_{\mathrm{eff}}}\;,
\eeq
where $k$ is Boltzmann's constant($1.38 \times 10^{-23}~J/K$) and $\eta\approx1$ is the system efficiency factor. The SEFD, tabulated in \tabref{SEFD}, can be regarded as the strength of a signal that, when coherently summed over all antennas, gives the same induced power as that of the noise. Recent measurements~\cite{Wijn11} support the general frequency dependence with an absolute value that is about 15\% higher 

\begin{table}[!ht]
 \caption{SEFD for a LOFAR core HBA antenna field consisting of 24 tiles. The last column gives the relative count rate for neutrino detection as function of frequency.}
 \tablab{SEFD}
 \begin{center}
 \begin{tabular}{|c||c|c|c|}
 \hline
 Freq & Core & $C_\nu$ \\
 \ [MHz] & \ [kJy] & \ [arb] \\
 \hline
 120 & 3.6 & 1.5 \\	
 150 & 2.8 & 1.0 \\	
 180 & 3.2 & 0.6 \\
 210 & 3.7 & 0.4 \\	
 \hline
 \end{tabular}
 \end{center}
 \end{table}

The optimum condition for the trigger is that the largest number of Moon pulses will be detected. For a given frequency $\nu$ we have calculated the relative count rate
\beq
C_\nu= \int dE\,\Phi(E) \, P_\nu(E) \;,
\eqlab{CntRate}
\eeq
where $E$ is the neutrino energy, $P_\nu(E)$ the chance of detecting a signal at frequency $\nu$ from a neutrino of energy $E$, and $\Phi(E)$ is the neutrino flux. The latter is often chosen proportional to $E^{-2}$~\cite{Wax98}. The detection probability is calculated using the procedure discussed in Ref.~\cite{Sch06} including a realistic frequency dependence of the pulse. The threshold for detecting a Lunar pulse is taken proportional to the SEFD given in \tabref{SEFD}, where the constant of proportionality cancels in taking ratios.

The relative count rates calculated from the SEFD and given in \tabref{SEFD} show that it is strongly favorable to include as many of the lower frequencies in the window as possible. Care should be taken with subbands number 100--200 where there are a large number of intermittent NRFI lines (see \figref{rfi_situation}). It should be noted that it is not advantageous to measure at even lower frequencies than given in the table partly because the smaller effective area of the LBA fields and partly because the rapid increase of the sky temperature.

\subsection{Frequency filter and pulse structure}\seclab{PulseStructure}

As mentioned before, due to bandwidth limitations only 244 of the 512 subbands can be sent to CEP for real-time processing. To make the optimum choice for this selection we have to take into account the considerations discussed above, i.e.\ lower frequencies give a larger aperture, and NRFI-free subbands should be selected that have a gain greater than half of the average. An additional consideration is that when the bandwidth limited signal is transformed back to the time domain, a pulse is still narrow in time such that a sensitive trigger can be constructed. The pulse form will depend on the particularities of the selection of 244 subbands, referred to as the Frequency-Filter Scheme (FFS)\footnote{We used 246 subbands in the simulation, as this used to be the maximum number of bands for the core. The difference will not affect our results.}. We have analyzed a few different FFSs.
For all choices we have omitted the low-gain as well as the NRFI corrupted subbands.
\begin{itemize}
\item[LoB] One large window at the lowest frequencies.
\item[Log] To give some weight to the higher frequency subbands the selected frequency channels follow a logarithmic pattern with a greater density at the lower frequencies. Including higher frequency components may sharpen the signal.
\item[Comb] As an extreme for sharpening the signal structure the frequency channels are selected in a comb-like structure of groups of 50 subbands which are NRFI-free.
\item[HiB] One large window at the highest frequencies. Even though this choice will not optimize the aperture, it diminishes the effects of ionospheric dispersion as will be discussed in later sections.
\end{itemize}

\begin{figure}[!ht]
  \centering
  {\includegraphics[width=0.5\textwidth]{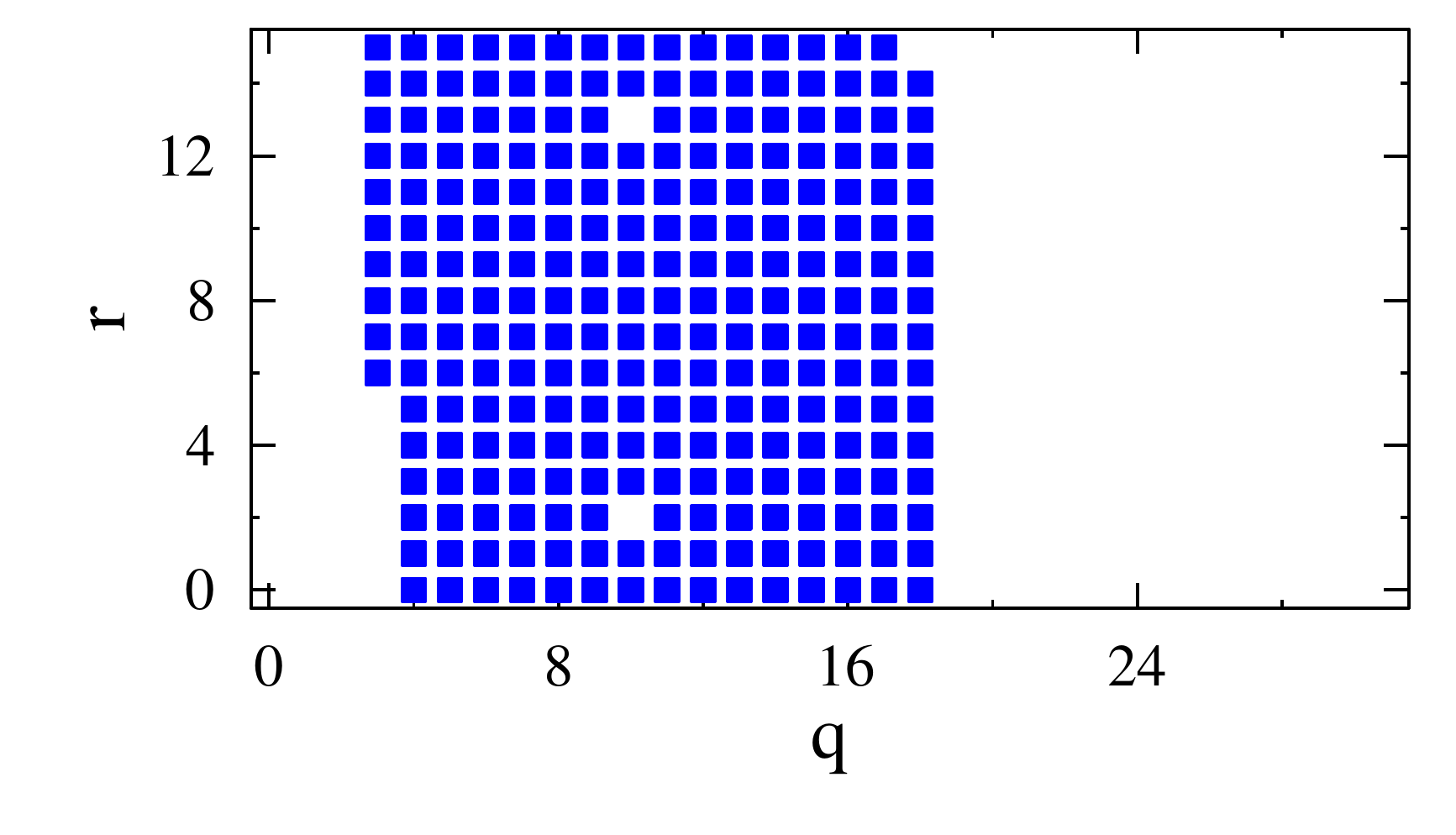} \
  \includegraphics[width=0.3\textwidth]{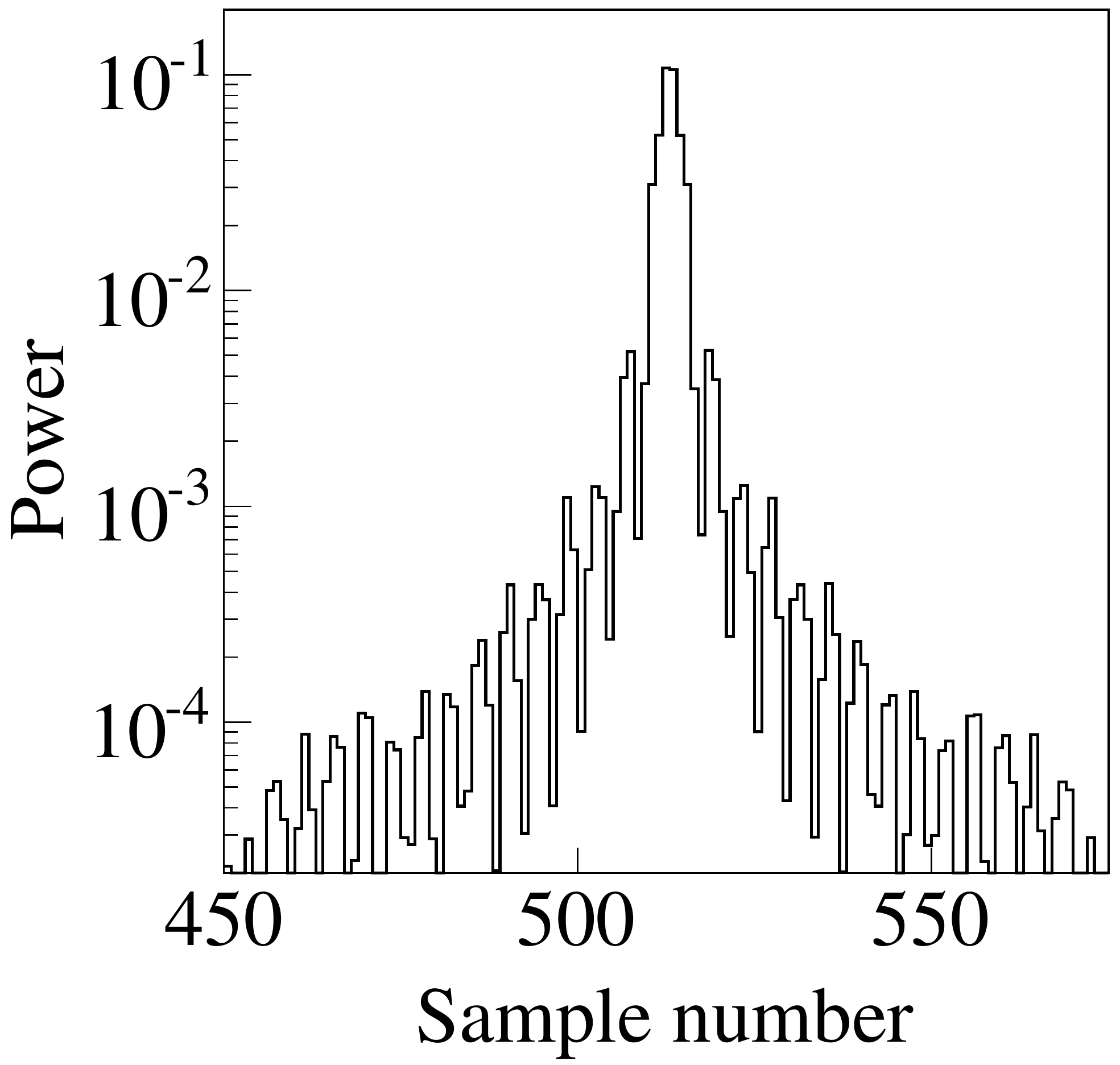}
  }
  \caption{For the LoB-FFS the selected frequencies are shown on the l.h.s.,\ while the r.h.s.\ shows the response of the filter to a very short bandwidth limited pulse.}
  \figlab{LoB_pulse}
\end{figure}

\begin{figure}[!ht]
  \centering
  {\includegraphics[width=0.5\textwidth]{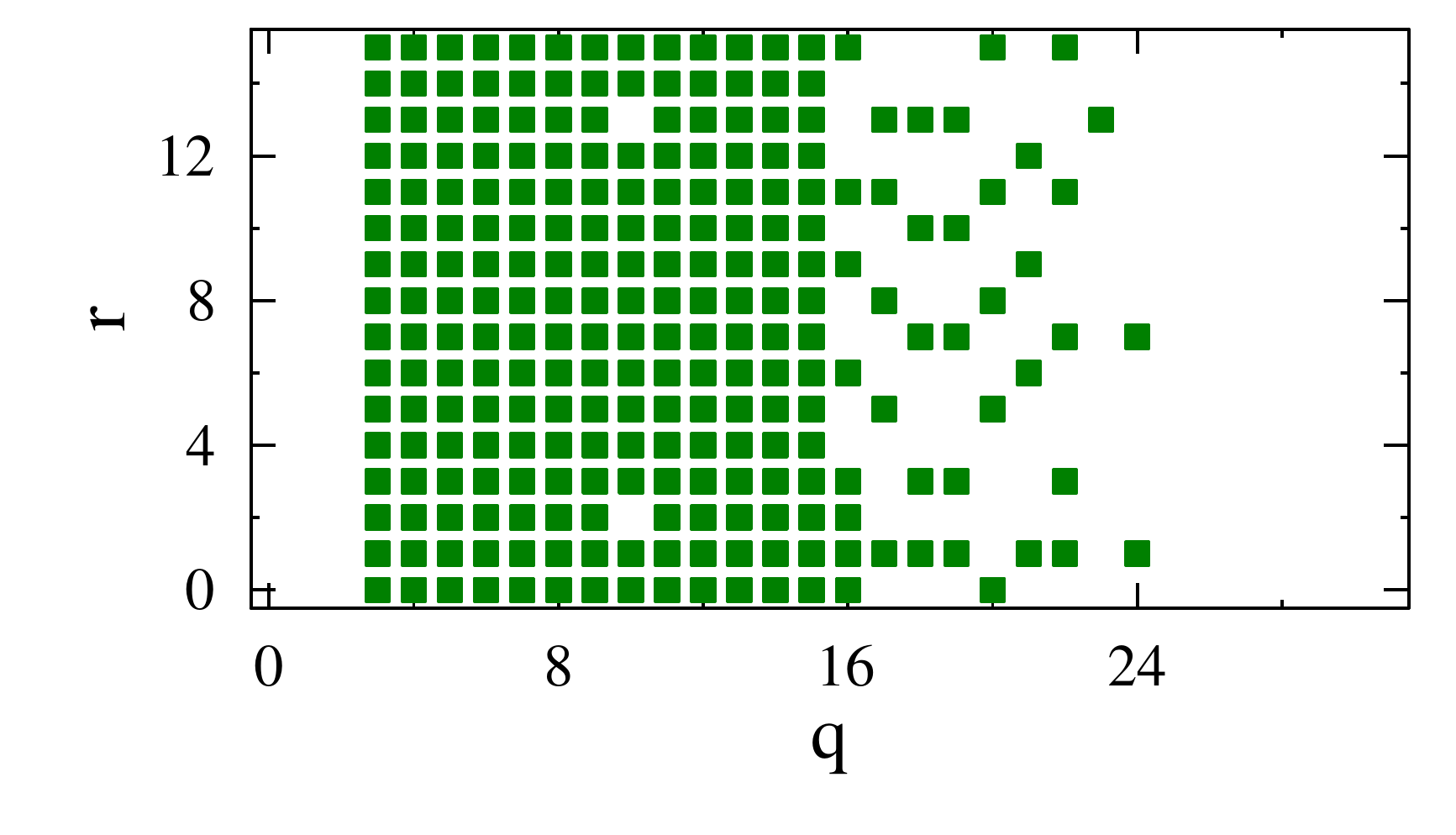} \
  \includegraphics[width=0.3\textwidth]{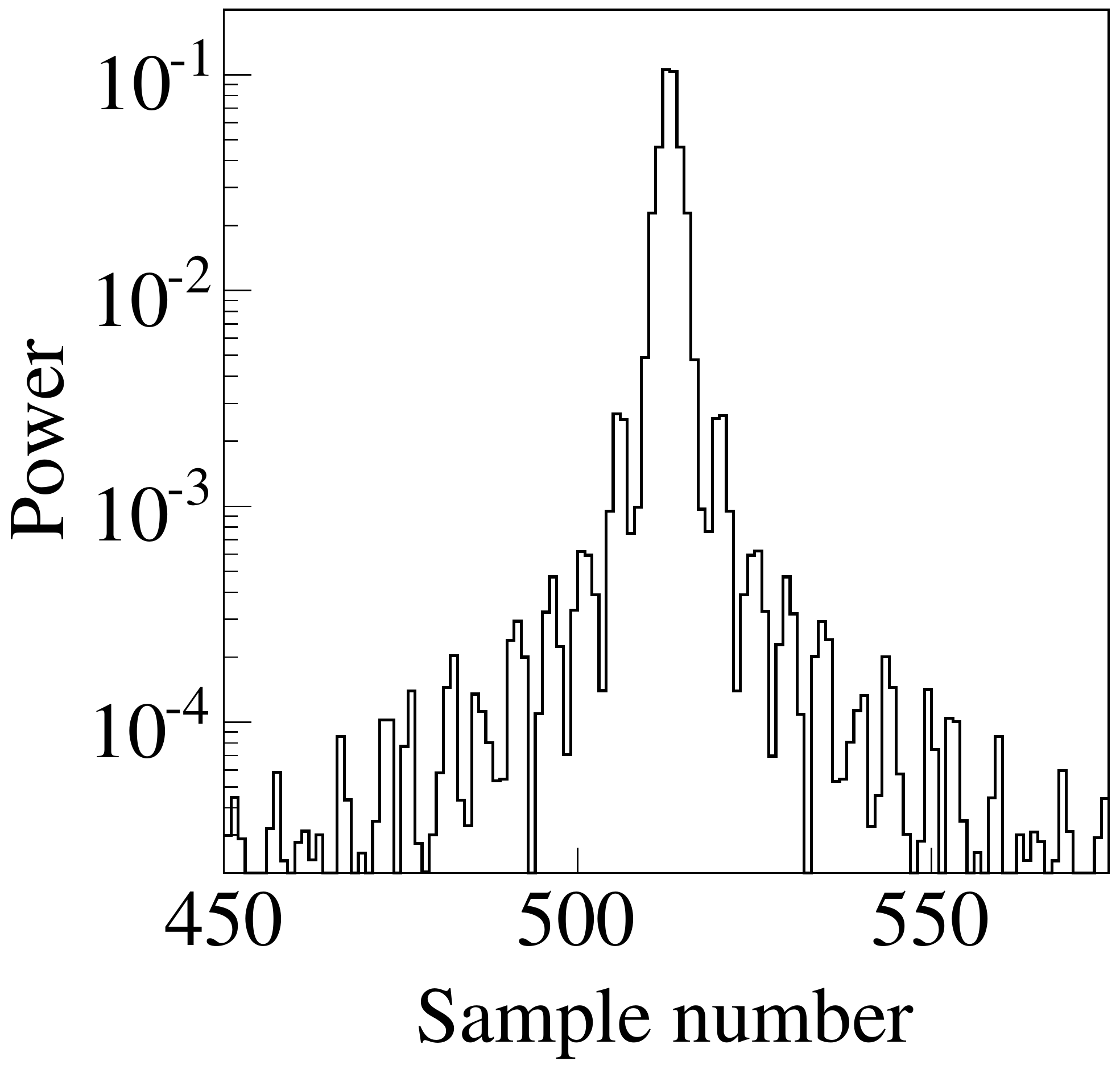}
  }
  \caption{Same as \figref{LoB_pulse} for the Log-FFS.}
  \figlab{Log_pulse}
\end{figure}

\begin{figure}[!ht]
  \centering
  {\includegraphics[width=0.5\textwidth]{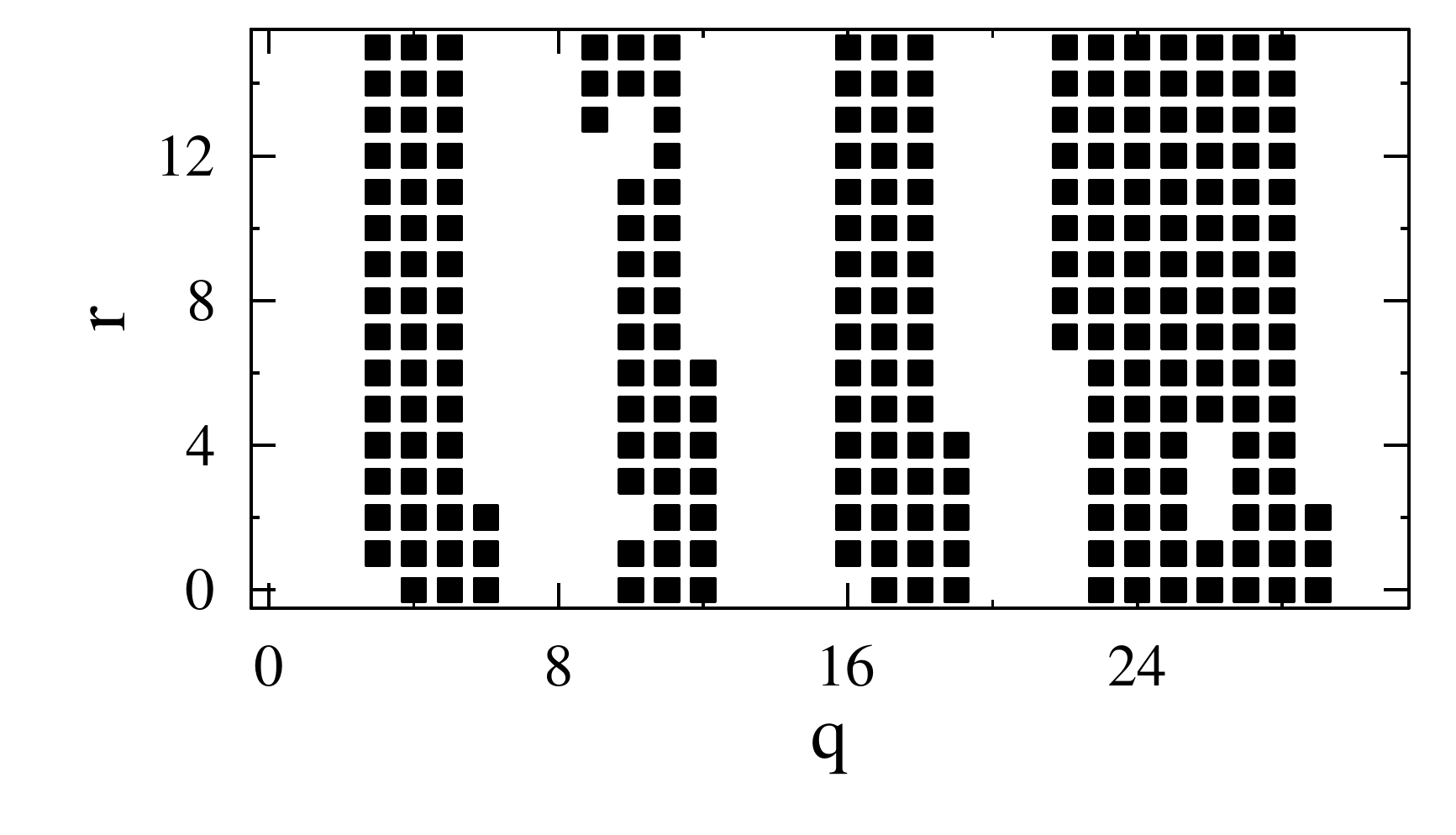} \
  \includegraphics[width=0.3\textwidth]{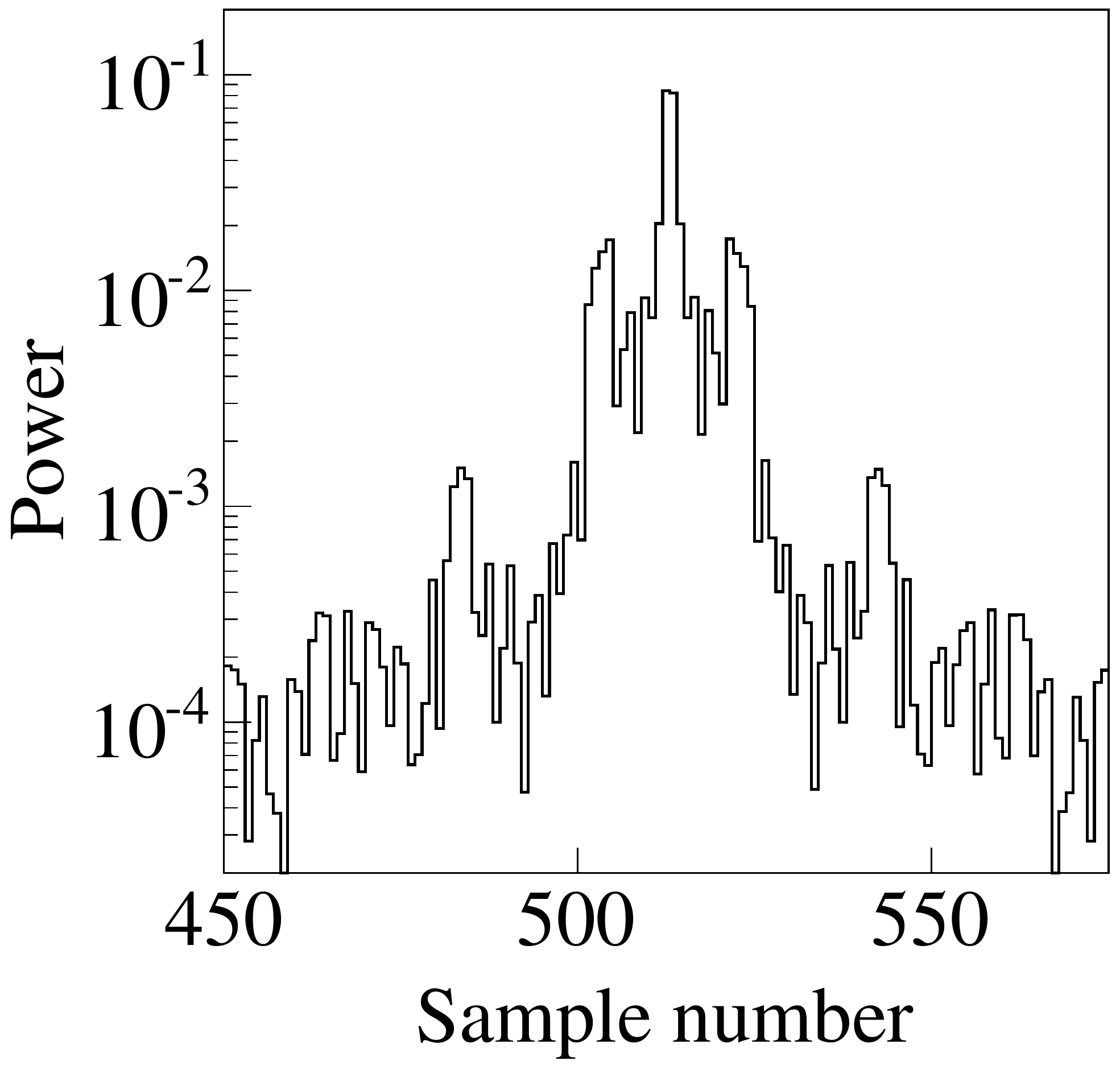}
  }
  \caption{Same as \figref{LoB_pulse} for the Comb-FFS.}
  \figlab{Comb_pulse}
\end{figure}

\begin{figure}[!ht]
  \centering
  \centering
  {\includegraphics[width=0.5\textwidth]{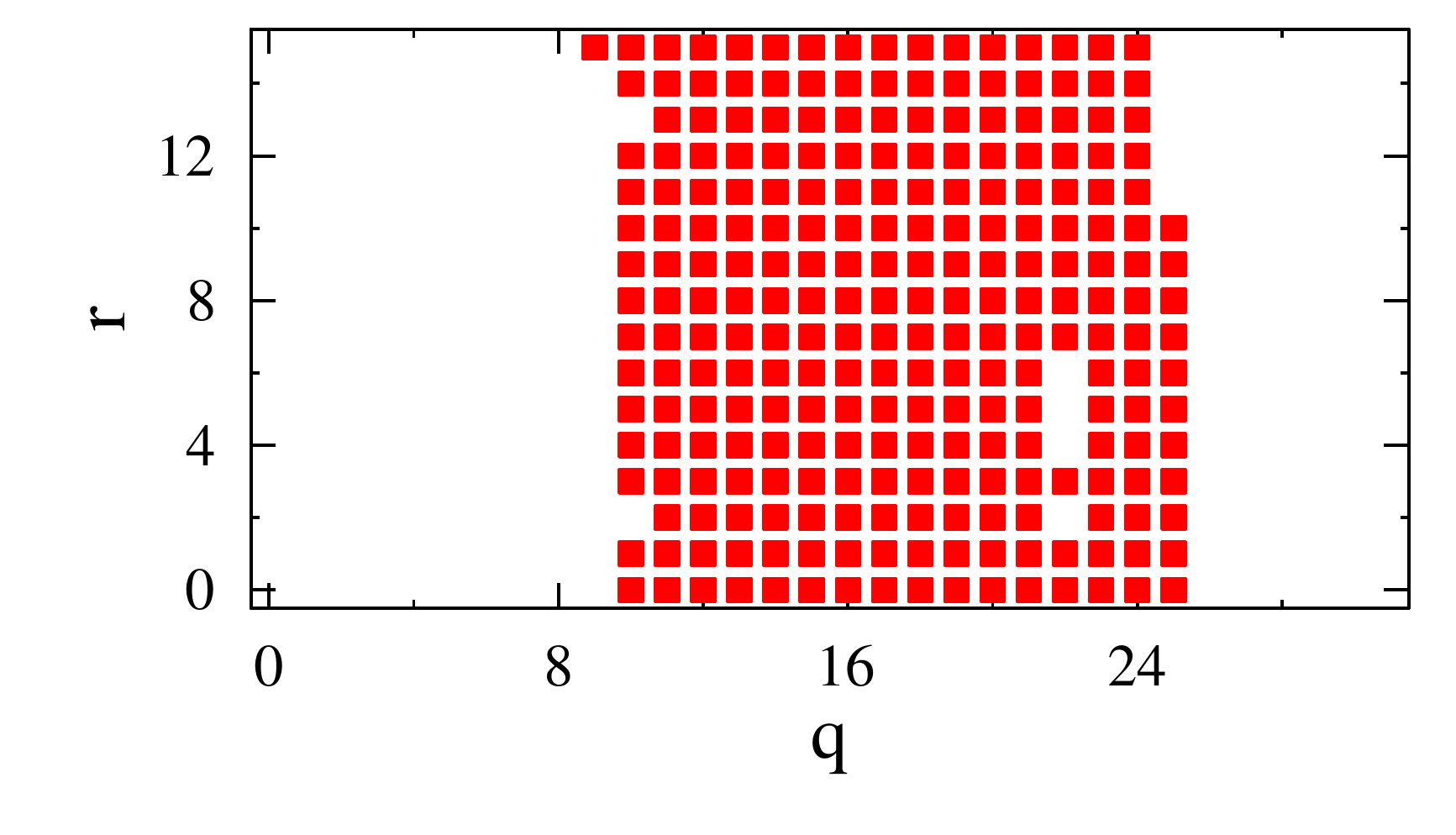} \
  \includegraphics[width=0.3\textwidth]{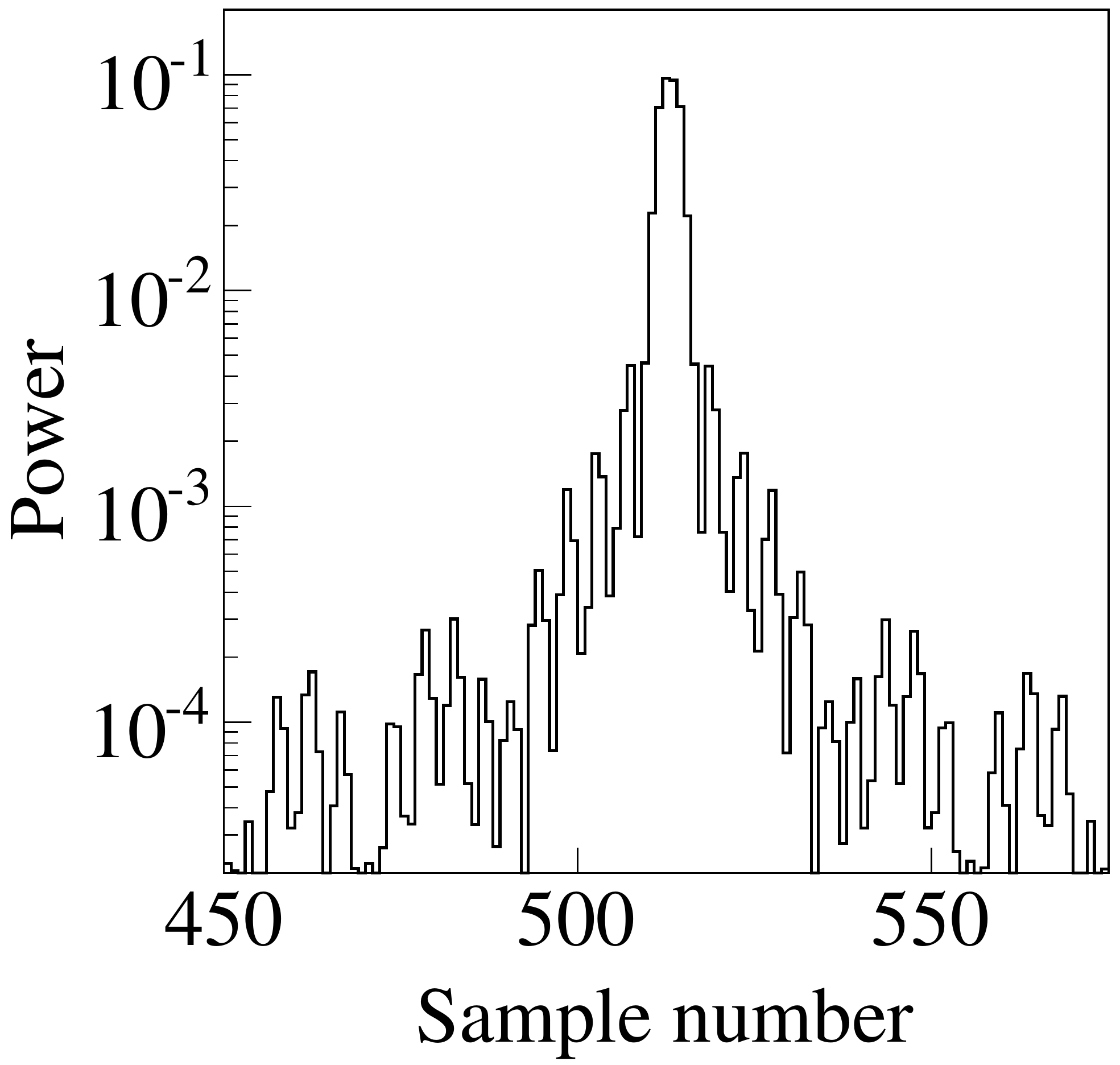}
  }
  \caption{Same as \figref{LoB_pulse} for the HiB-FFS.}
  \figlab{HiB_pulse}
\end{figure}

Each of these FFSs is illustrated in Figures \ref{fig:LoB_pulse}--\ref{fig:HiB_pulse}. The selected frequency window is shown on the l.h.s., where the subband number is equal to $16q+r$. The r.h.s.\ displays the corresponding pulse response in units where the original pulse carries unit power. It is clear that the pulse response is very different for the various FFSs which will be reflected in the efficiency of recovering it from the noise.
The additional effects of ionospheric dispersion will be investigated in \secref{IonDis}.

\subsection{Noise with different filtering methods}\seclab{RealData}

\begin{figure}[!ht]
 \centering
 \includegraphics[width=.5\textwidth]{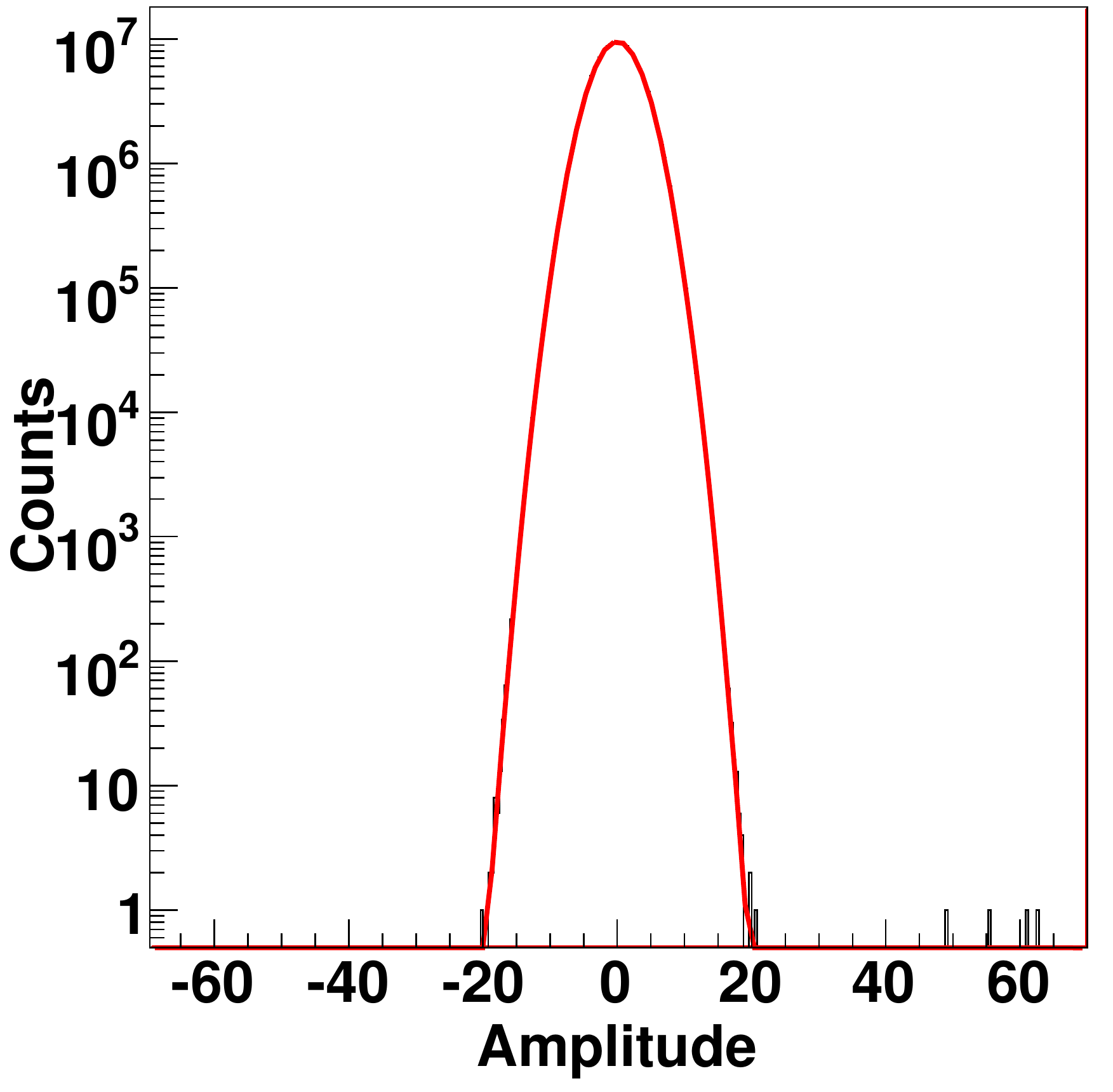} 
 \caption{The number of times an amplitude is observed in 1\,s of data stream of single HBA tile of LOFAR after NRFI mitigation is plotted vs.\ amplitude for a unit bin-size. The smooth curve (hardly distinguishable from the histogram) shows a fitted gaussian to the data.}
 \figlab{ampl_histogram_full}
\end{figure}


\Omit{
Filter                sigma of the
              fitted Gaussian profile         3sigma                5sigma           7sigma
-------------------------------------------------------------------------------------------
Comb                     2.552             2.95397 X 10$^5$           317              104
HiB                      2.848             2.57694 X 10$^5$           196               50
LOG                      2.020             2.84089 X 10$^5$           284              135
LoB                      2.197             2.87584 X 10$^5$           347              178
}

Before adding a pulse to a background (noise) spectrum we investigate the structure of the data, in particular the extent to which the noise can be regarded as Gaussian.
For this we have processed 1 second raw time-series data from a single HBA tile of LOFAR.
These data contain no NuMoon pulses; it is simply a sample of the noise levels of LOFAR.  The data are passed through the simulated PPF (see Appendix A), after which the NRFI lines in frequency domain are removed (see \secref{RFIMitigation}). The data are transformed back to the time domain by applying the PPF inversion routine. The resulting amplitude distribution is shown in \figref{ampl_histogram_full}. The drawn curve shows a Gaussian profile fitted to the data. The $\chi^2$ of the fit is close to unity showing that the noise closely resembles Gaussian noise.
Due to noise transients, the data shows a small number of large pulses well above the expectation based on the Gaussian profile.
A closer investigation of these large pulses indicates that they are single timing-sample upsets.
On the basis of the experience obtained from observations with WSRT~\cite{Sch09NM,Bui10} and preliminary analysis of LOFAR data we expect that most of them diminish in importance when the signals of a large number of antennas are coherently added and the that remaining ones can be eliminated by the requirement that they originate from a well-defined spot on the Lunar surface.

On the basis of these results we conclude that for an investigation of the relative merits of the various FFSs it is sufficient to run simulations where a pulse is added to a Gaussian-noise spectrum.

\section{Pulse-search algorithm}\seclab{PulseSearch}

Central to the trigger algorithm is the pulse-search routine. For this we investigate the most efficient way to search the data for short, bandwidth-limited pulses of the type that may result from a cosmic ray or neutrino hitting the Moon. The basic search algorithm consists of measuring the power of an incoming signal over a certain amount of time.  This can be visualized as a window of time sliding over the data.  We must identify both an optimum FFS and an optimum size $N$ for the window sliding over the data.  This is done through  simulations where we add  pulses of different magnitudes to a spectrum of simulated pure Gaussian noise. The magnitude of these pulses is measured in terms of the average noise power, $\sigma^2$.  A pulse is added at a random time-position in every third page of a set of 3 pages. Each page contains 1024 time-samples.

Our aim is to design triggering criteria such that a large percentage of pulses from the Moon will be processed while suppressing random noise triggers. For definiteness we have set the random-trigger level at about once every minute.  In realistic situations one is limited by the system's dead-time (estimated at about 5 seconds per event) and storage capabilities. The deadtime is inherent to the way LOFAR manages data:  LOFAR's TBBs and station processors use the same data buses to communicate with CEP.  Triggering causes the TBBs to dump data to CEP, and while this dumping is in progress no new data can be recorded at the TBBs.  While not long, this deadtime will cut into the efficiency of NuMoon if triggering occurs too frequently.  As mentioned earlier, the use of anti-coincidence criteria will reduce the number of triggers caused by transient noise. In this work we apply the pulse-search algorithm separately to each polarization.
In the calculation of the sensitivities to pulses from UHE neutrinos, this is accounted for by assuming that the pulse power is distributed 50-50 over the two polarizations. This will constitute an underestimate of the efficiency since adding the two polarizations incoherently will increase the signal over noise ratio.

\subsection{Power of $N$ consecutive time samples ($P_{N}$) }\seclab{PN}

To analyze the time series we retrieve the power from a sliding window of size $N$ bins of 5\,ns,
\beq
{P_{N}}(i)= {1\over \sigma^2} \sum_{n=1}^{N} v^2_{(i+n)} \;.\eqlab{PN}
\eeq
where $v_i$ is the voltage for the $i^{th}$ time sample. As mentioned before, the noise power $\sigma^2$ is defined as the average power per time sample for a full bandwidth spectrum, after subtracting the sharp-frequency RFI lines.

For every page of 1024 time samples, the maximum power in the window is defined as
\beq
 P^\mathrm{m}_N=\max_i P_N(i) \;, \eqlab{PmN}
\eeq
Depending on the value of this maximum, a trigger flag will be set.  In order to choose a threshold for $P^\mathrm{m}_N$ we first analyze the structure of the noise which depends on the FFS that is used.

\subsection{Accidental noise pulses \& threshold determination}\seclab{Threshold}

Sometimes noise will cause a trigger-flag to be set. This is referred to as an accidental trigger.  In order to predict the rate at which accidental triggers occur, we have analyzed the noise with the sliding window method. The distribution of $P^\mathrm{m}_N$ values is determined for a sample of 1~second filled with Gaussian noise. This analysis is repeated for all FFSs and for a range of window sizes $N$. The general features of the distribution are independent of the particular FFS or the value of $N$ that has been used. As an example, in \figref{noise_filter} the distribution of $P^\mathrm{m}_N$ is plotted for $N=7$ and the LoB-FFS.

\begin{figure}[!ht]
 \centering
 \includegraphics[width=.4\textwidth]{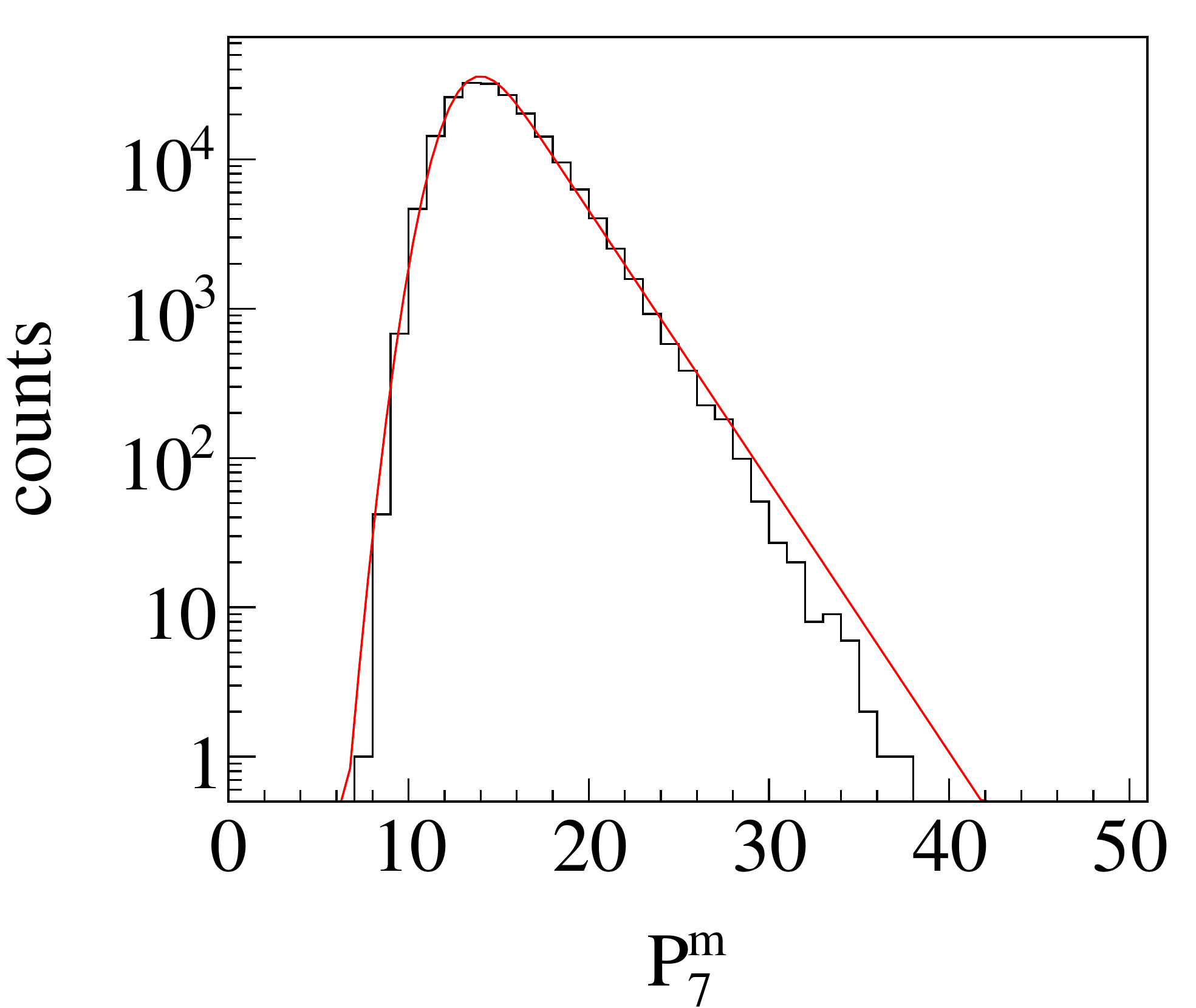}
 \caption{Plotted are the number of occurrences of a $P^\mathrm{m}_N$ value per unit bin size for 1~second of Gaussian noise filtered with the LoB-FFS for $N=7$. The drawn, red, curve shows the fit to the spectrum using \eqref{Fx}.}
 \figlab{noise_filter}
\end{figure}

In the analysis we aim to set a threshold value for $P^\mathrm{m}_N$ which will result in a certain maximal accidental trigger rate. For the present analysis, we have set the maximal accidental trigger rate at once per minute. For each $P^\mathrm{m}_N$ distribution, we have determined the threshold value $P^\mathrm{t}_N$ where we expect to find a value  $P^\mathrm{m}_N > P^t_N$ only once per minute. The value of the threshold is determined by fitting a particular function $F(x)$ to the distribution of $P^\mathrm{m}_N=x$.
For a large value the distribution should follow that of a $\chi^2$ distribution with $k$ degrees of freedom,
\beq
P(x,k) \propto x^{k/2-1} e^{-x/2}  \;. \eqlab{Ptk}
\eeq
To a good approximation the number of independent degrees of freedom in the distribution is given by $k=N/2$ since, due to the FFS, the signal is oversampled by almost a factor two. For simplicity we have chosen to fit the spectrum by a convolution of a Gaussian and an exponential that is cut off at the lower end,
\begin{equation}
F(x) =  B\, \int_{p_{co}}^{\infty} e^{-(x-x')^2/\sigma^2} \, e^{a\, x'} \, \mathrm{d}x'\; , \eqlab{Fx}
\end{equation}
with fitting parameters $B$ (a normalization factor), $\sigma$ (the width of the Gaussian), $p_{co}$ (the x-value where the exponential is cut off) and $a$ (the slope of the exponential). $F(x)$ is integrated to determine the value of $P^\mathrm{t}_N$ which will correspond to the desired accidental trigger rate. As is clear from \figref{noise_filter}, the distribution is overestimated for large values of $P^\mathrm{m}_N$. Working with the fitted curve thus gives rise to a higher value for $P^\mathrm{t}_N$ than would be necessary on the basis of pure Gaussian noise.

The thus determined values of $P^\mathrm{t}_N$ for the various FFSs and window-sizes are given in \figref{thresholds}. With increasing window-size $N$ one expects the threshold $P^\mathrm{t}_N$ to increase, since the time-average power in a window is proportional to the size of the window.  This explains the rising trend one sees in the determined values of $P^\mathrm{t}_N$ for each FFS. Note that the various FFSs introduce correlations in the noise-spectra, which are reflected in the differences one sees in their $P^\mathrm{t}_N$ values.

\begin{figure}[!ht]
 \centering
 \includegraphics[width=.5\textwidth]{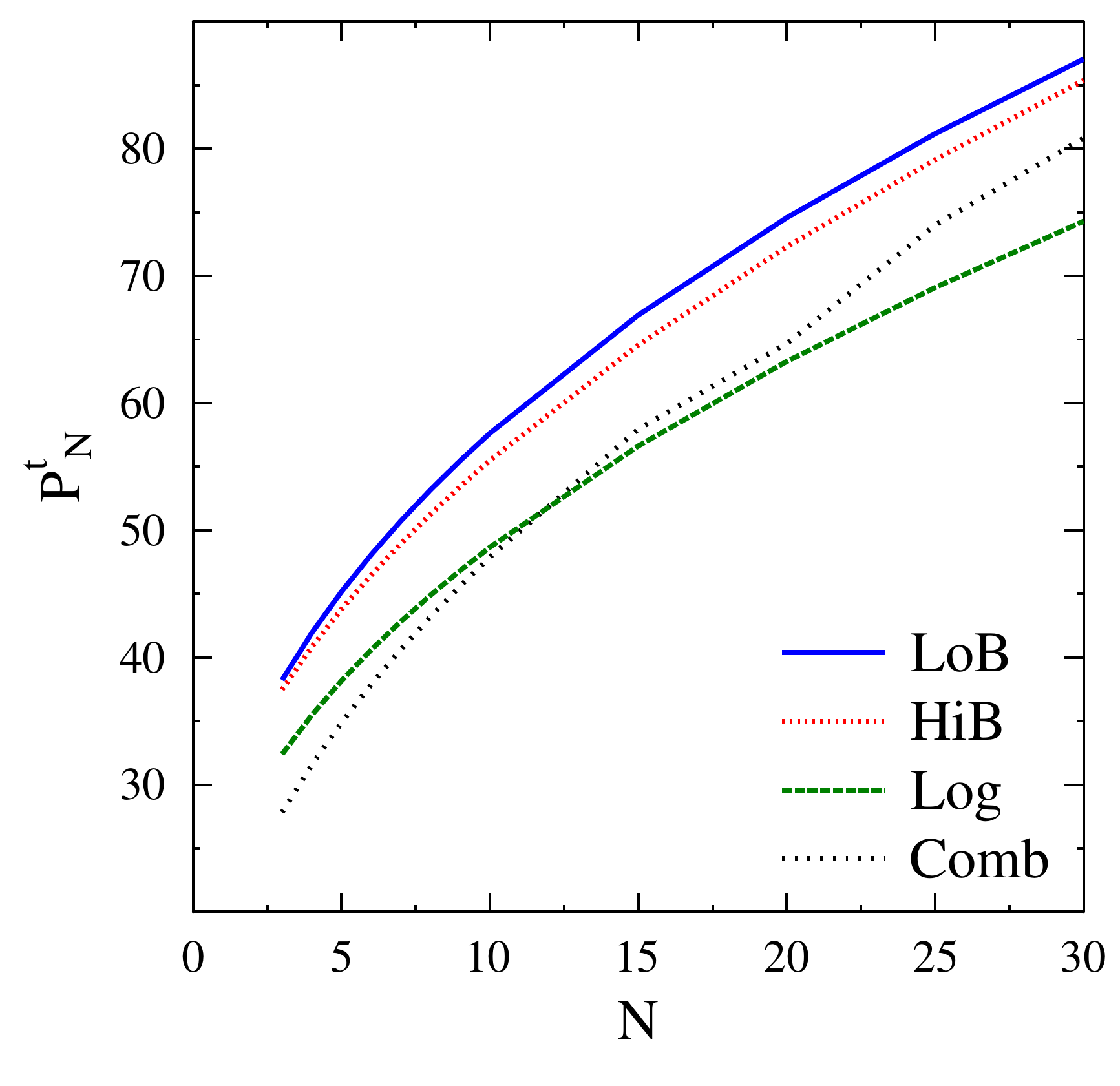}
 \caption{The determined threshold values $P^\mathrm{t}_N$, corresponding to 1 accidental count per minute, as function of $N$ for the different FFS under consideration.}
 \figlab{thresholds}
\end{figure}

\subsection{Pulse amplitude distribution}\seclab{PulseAmpl}

As was shown in \secref{PulseStructure}, the selected FFS strongly affects the measured shape of the pulse in time, and in general the pulse will broaden. With increasing size ($N$) of the sliding window, a larger fraction of the broadened pulse will be recovered. However, increasing the size of the sliding window will also capture more noise power (see \figref{thresholds}). At a certain point, this captured noise power will no longer balance the increase in captured pulse power causing a worsening of the signal-to-background ratio. As a first step towards finding the optimum size of the window, we investigate how well the original power of the pulse is recovered by the sliding-window procedure.

We start with a very short, delta-function like in time, pulse of unit power (when integrated over the full bandwidth of the HBA subbands, before applying and filters) placed at a random position in a page. The pulse is processed as described earlier: The FFS is applied after a 16 tap PPF, and then the signal is converted back into the time domain by applying a full PPF-inversion (PPF$^{-1}$). The maximum power found in a window of length $N$ is taken to be the recovered power of the pulse.  This recovered power is compared to the original power of the pulse (here, original power = 1). The recovered power depends on the structure of the recovered pulse, which in turn depends on the phase of the sampling-cycle at the time the pulse arrives. (One full sampling-cycle is equal to one time-sample.) To account for the fact that a pulse may arrive at any phase of the sampling-cycle, the analysis is repeated 1000 times with the pulse arriving at various phases of the sampling-cycle.
The average value of the recovered power, $\bar{P}^m_N$, is shown as a percentage of the power of the input pulse in \figref{RetrPower}. This procedure was applied for every value of $N$ between 3 and 50, and for each of the FFSs. This part of the analysis has been done without adding noise to the spectrum.

\begin{figure}[!ht]
 \centering
 \includegraphics[width=.5\textwidth]{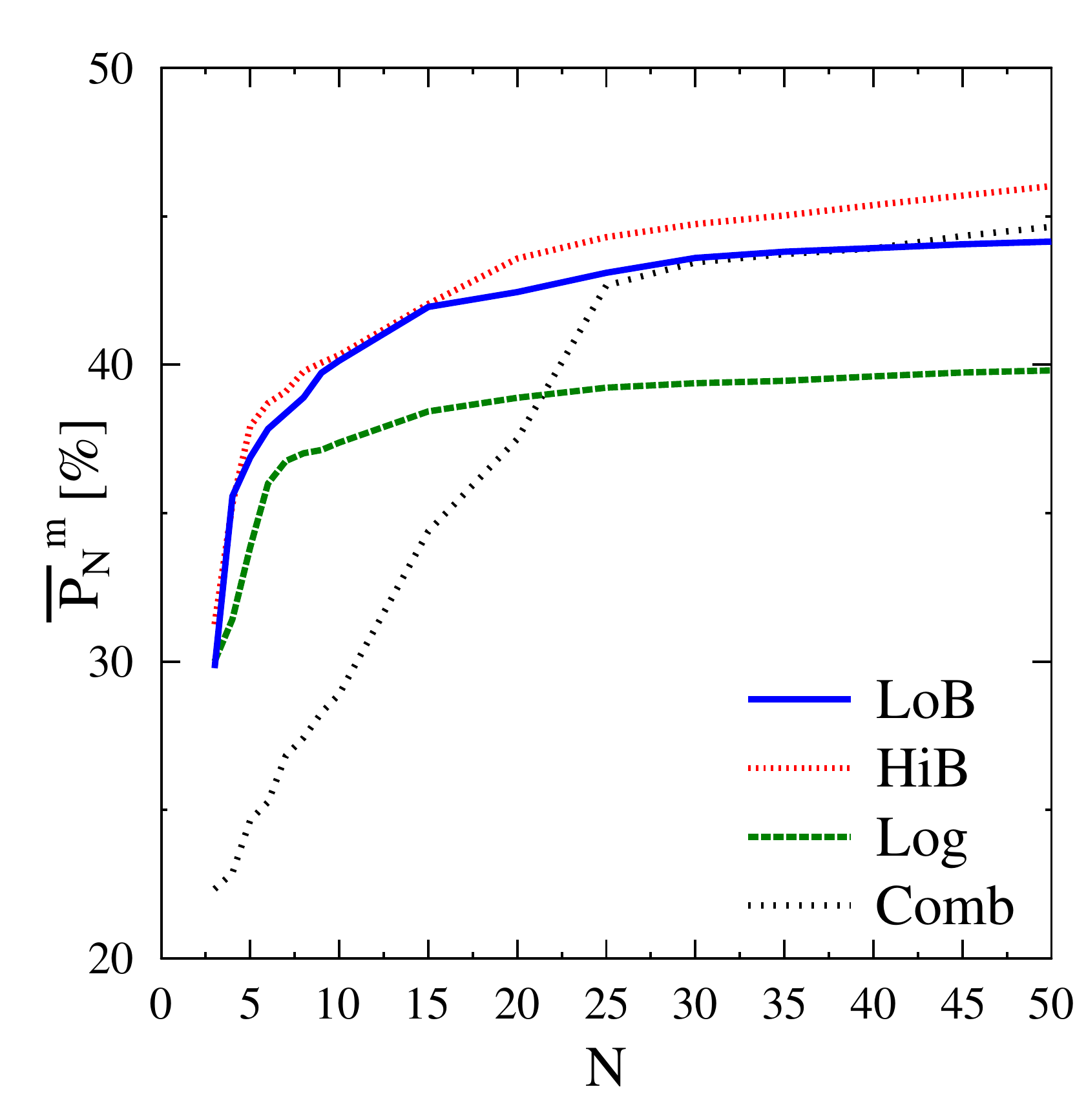}
 \caption{Percentage of retrieved power for pulses with random phases for the different FFSs. This analysis is performed without a noise background.}
 \figlab{RetrPower}
\end{figure}


With increasing window length $N$, an increasing fraction of the power of the input pulse is recovered. The retrieved power saturates at about 50\% due to the bandwidth of the FFSs. For small values of $N$ the Comb-FFS performs considerably worse than the other FFSs. This can be understood from \figref{Comb_pulse} where it is shown that the peak of the power distribution is considerably wider than that for the other FFSs. The lower saturation value for the Log-FFS is due to the fact that in this FFS there is considerably more power in channels more than 100 time samples removed from the peak (outside the range shown in figures (\ref{fig:LoB_pulse} $\cdots$ \ref{fig:HiB_pulse}).
As is shown in \secref{Threshold} the noise will continue to increase with $N$. We expect that there is an optimum for $N$.  To search for this optimum $N$, we repeat the previous analysis with noise included.

\begin{figure}[!ht]
  \centering
  \includegraphics[width=0.6\textwidth]{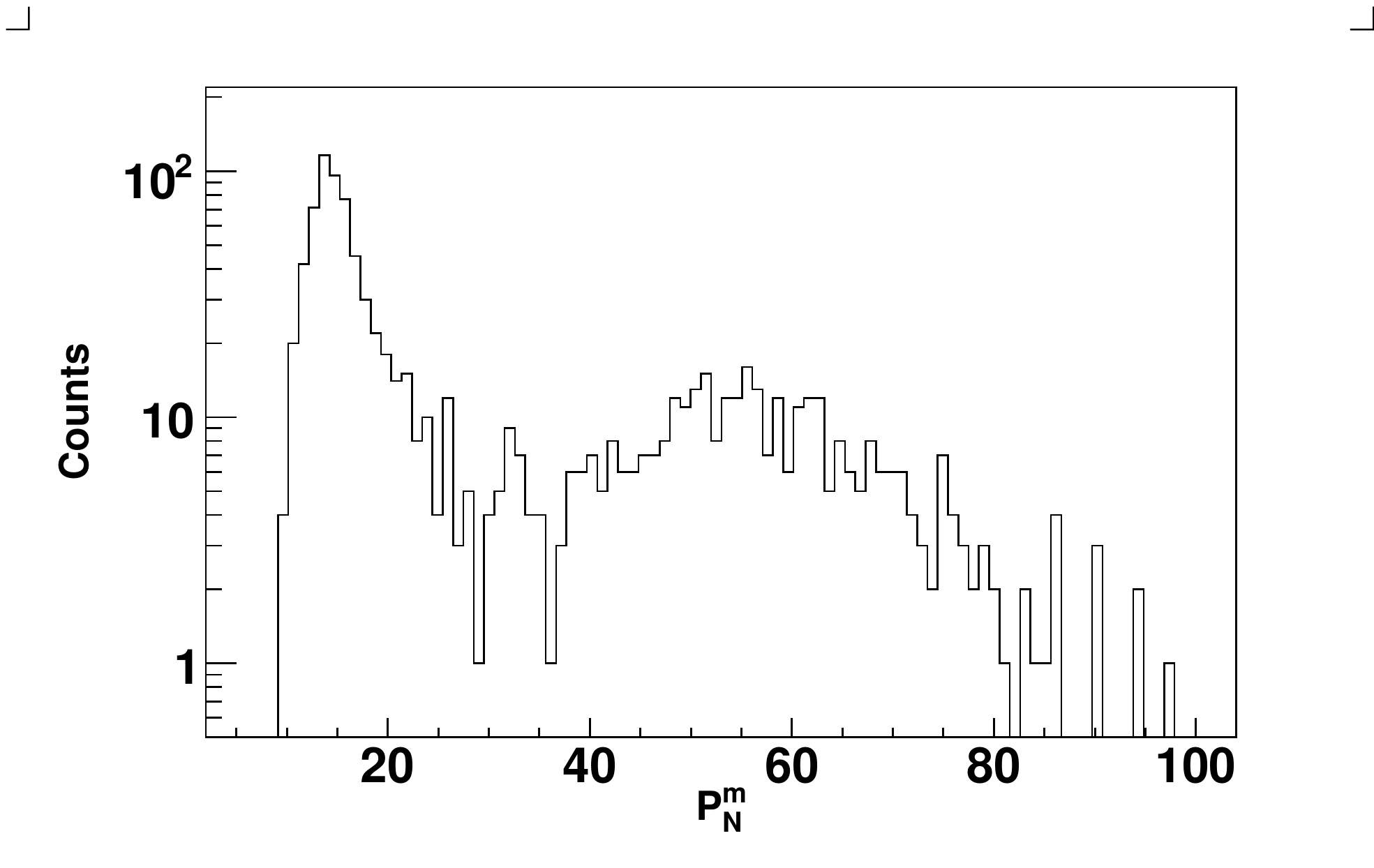} 
  \caption{Histogram of the number of occurrences of a $P^\mathrm{m}_N$ value per unit bin size when a pulse, with power 144$\sigma^2$, is added to a noisy background for every third page and filtered with the Comb-FFS, using $N=15$.}
  \figlab{power_spectrum1}
\end{figure}

When including noise in the analysis the picture becomes more complicated since the pulse and the noise will interfere. To study this case we have analyzed 1000 pages of 1024 time-samples each containing Gaussian noise with a power of $\sigma^2$ per time sample. To each third page a pulse with a predetermined power is added with a random phase at a random position. The time traces are run through the complete simulated NuMoon pipeline, including the FFS, and for each page the value of $P^\mathrm{m}_N$ is determined.
As an example, the spectrum of $P^\mathrm{m}_N$ values for $N=15$ is plotted in \figref{power_spectrum1} for a pulse with power 144\,$\sigma^2$ using the Comb-FFS.
At lower values of the power, the noise is following the spectrum shown in \figref{noise_filter}, since two-third of the analyzed pages contain exclusively Gaussian noise. Centered around a value of about 60 a broad bump shows. This is due to those pages where a pulse was added.

It is instructive to develop a feeling for the numbers. When a pulse of power $A^2\times \sigma^2$ is added to noise this will give a broad structure in the spectrum of $P^\mathrm{m}_N$ values with the centroid at $\bar{P}^m_N=A^2\times E +N/2$ where $E\approx 0.4$ is the efficiency of power reconstruction (see \figref{RetrPower}). Since the FFS approximately halves the bandwidth, a window of length $N$ contains a noise power of $\sigma^2N/2$. Due to interference with the noise the structure extents from $P^\mathrm{m}_N=(A-1)^2\times E +N/2$ till $P^\mathrm{m}_N=(A+1)^2\times E +N/2$ and has thus a width of $\Delta P^m_N=4 A\times E$. The value of the threshold for which about 80\%  of the added pulses is recovered thus can be approximated as
\beq
P^t \approx (S_{80}-1)^2 \times E + N/2
\eeq
or inverted as
\beq
S_{80} \approx \sqrt{P^t/E}+1 \eqlab{S80Pt},
\eeq
essentially independent of window length $N$. The polarization degrees of freedom have not been considered.
For $N=15$, $A=12$ and $E=0.35$ one thus expects on average a value $\bar{P}^m_N=144 \times 0.35+N/2=58$, which agrees well with the result shown in \figref{power_spectrum1}. Also, the predicted width of the structure $\Delta P^m_N=4 A \times 0.35=17$, which agrees with the figure.
On the basis of these consideration, for a 1 min observation the $S_{80}$ value for $P_N^t=58$ is thus expected to be $S_{80}^{1m}=\sqrt{58/0.35}+1=13.8$, which is close to the value given in \figref{de_TimeBins} for the Comb-FFS with $N=15$.

For the data analysis it is important to know what percentage of added pulses of a certain magnitude produces a value for $P^\mathrm{m}_N$ that exceeds the trigger threshold $P^\mathrm{t}_N$, which was discussed in \secref{Threshold}. This number, the detection efficiency, is discussed in the following section.

\subsection{Detection efficiency}\seclab{DetEff}

To quantitatively compare the different FFSs and window-sizes, we have added pulses to Gaussian noise at random positions in every third data page of 1024 time-samples each, as discussed in the previous section.
The data 
are run through the complete signal processing chain (see \figref{flowchart}) including the PPF transformation, beam forming, selecting NRFI-free frequencies and the back transformation to time sampled spectra.
A trigger-flag is set when the value of $P^\mathrm{m}_N$ for one data page exceeds the threshold value $P^\mathrm{t}_N$ (discussed in \secref{Threshold}). The detection efficiency (DE) for a particular combination of FFS and $N$ is defined as the fraction of added pulses that generate a trigger signal.

\begin{figure}[!ht]
 \centering
 \includegraphics[width=.5\textwidth]{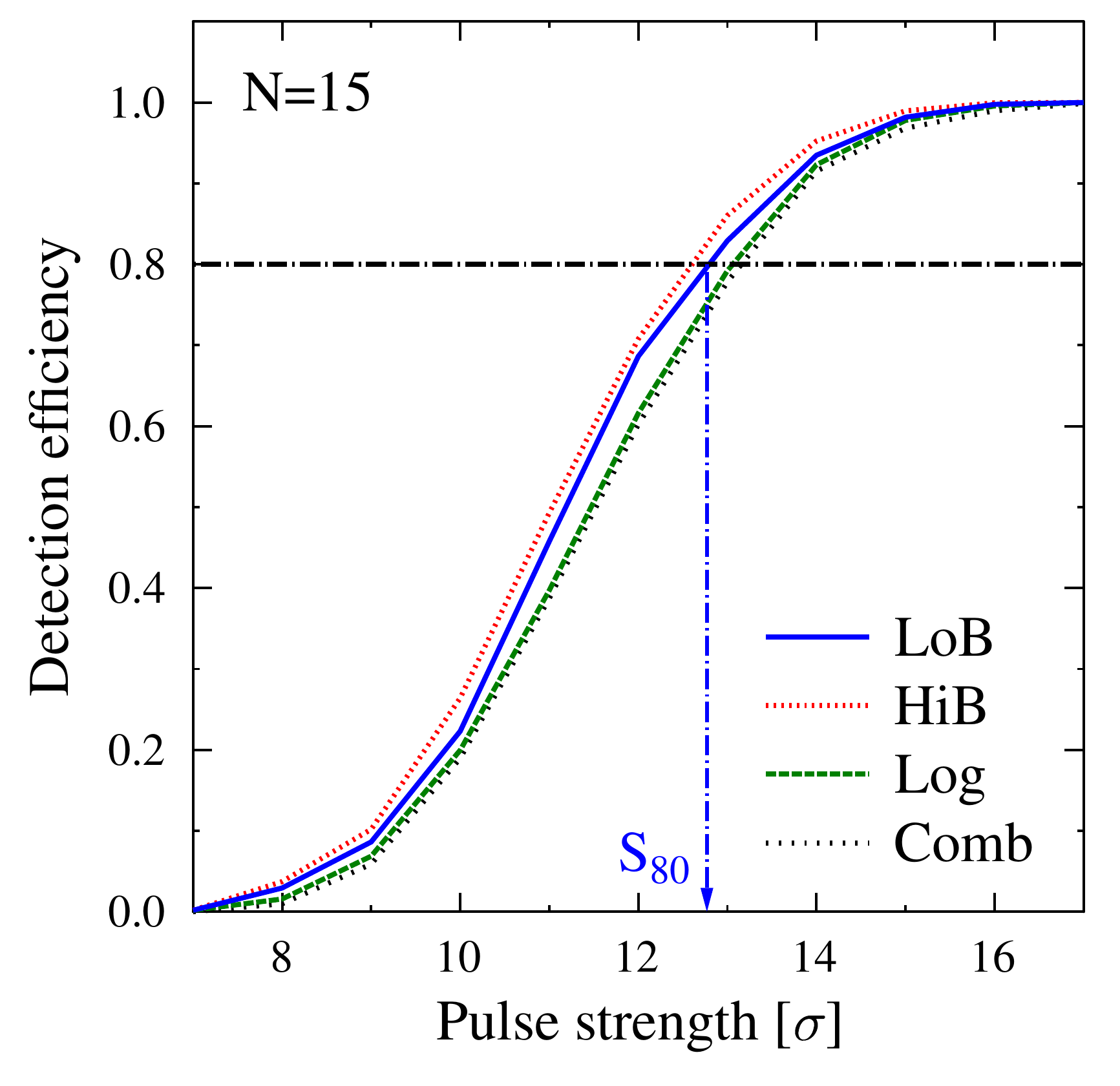}
 \caption{Detection efficiency is analyzed 
for the different FFSs 
with N=15 as a function of the power of the pulse. The 80\% recovery limit is indicated by the dash-dotted curve.}
 \figlab{DE-dTEC=0-dza=00}
\end{figure}


In \figref{DE-dTEC=0-dza=00} the DE is compared for the various FFSs. The DE is given as a function of the strength of the added pulses. Similar plots are made for a range of sizes $N$ of the sliding window. For a good operation of the NuMoon trigger scheme we demand a DE of 80\% or better. For each combination of FFS and $N$, we can determine a pulse-amplitude $S$, in units of $\sigma$, for which the detection efficiency is 80\% ($S_{80}(N)$). For each FFS the value of $S_{80}$ is plotted as function of $N$ in \figref{de_TimeBins}.

\begin{figure}[!ht]
 \centering
 \includegraphics[width=.5\textwidth]{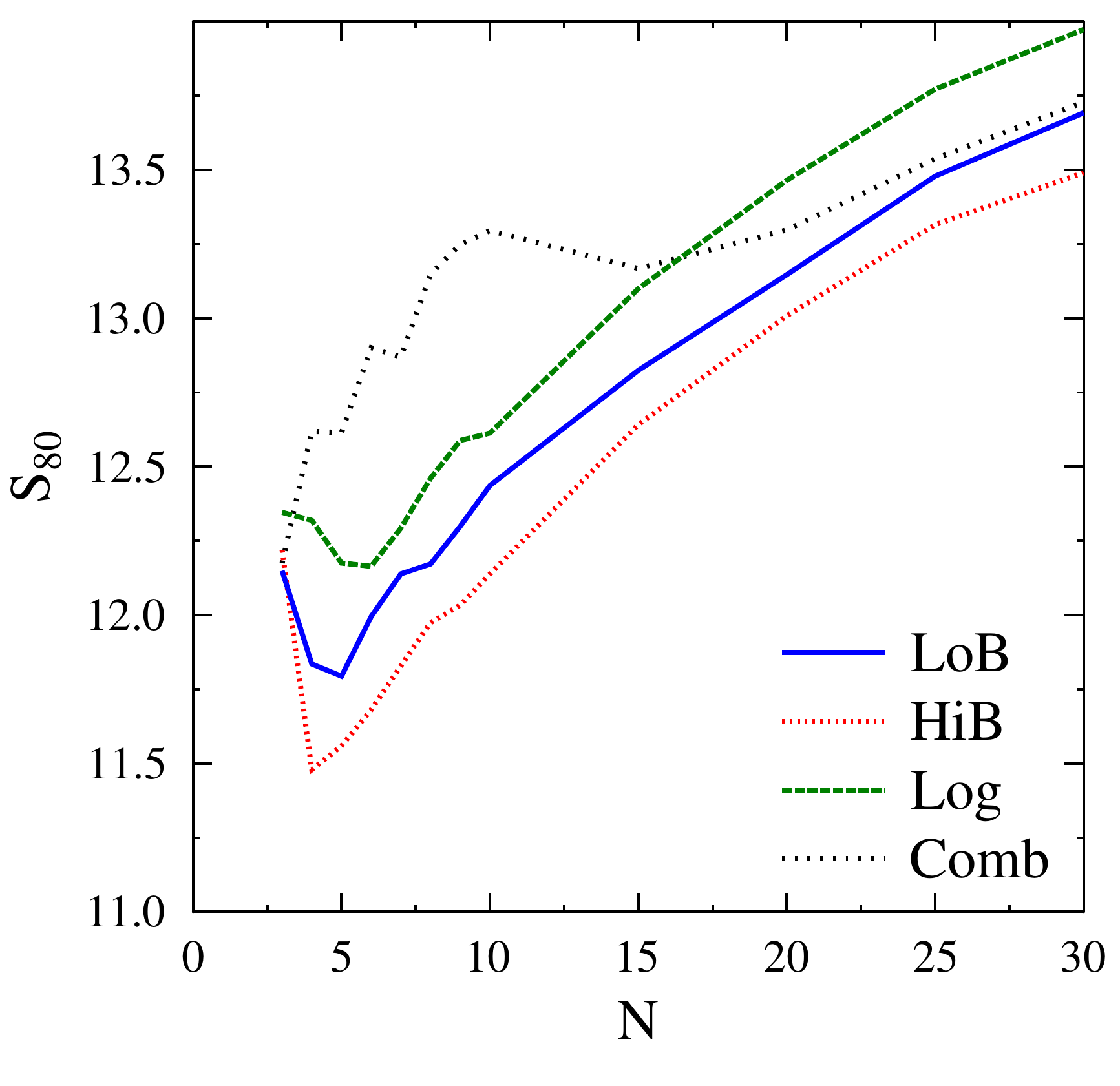}
 \caption{The pulse strength is given for which the DE exceeds 80\% ($S_{80}$) as function of the window size $N$ for the different FFS under consideration.}
 \figlab{de_TimeBins}
\end{figure}

For most of the FFSs, \figref{de_TimeBins} shows a trend 
that we can easily explain based on the previous discussions. With increasing window length the noise power in the window increases, which necessitates an increase in threshold value, $P^\mathrm{t}_N$, to reach a constant accidental count rate (see \figref{thresholds}).
The value of $P^\mathrm{t}_N$ increases faster than the recovered pulse power, shown in \figref{RetrPower}, resulting in increasing values for $S_{80}$ which are seen in \figref{de_TimeBins} at large values of $N$.
For all FFSs, except for the Comb-FFS, the initial decrease in  $S_{80}$ for small values of $N$ is thus clearly due to the strong increase of pulse power.
For the Comb-FFS the recovered pulse power increases step-wise because of the satellite structure of the pulse induced by this particular FFS (see \figref{Comb_pulse}) 
 and the drop in $S_{80}$ is seen only around $N=15$ where the first satellite starts to fall inside the sampling window.

\section{Ionospheric dispersion}\seclab{IonDis}

Since we are looking for short radio pulses coming from the Moon, we need to correct for the ionospheric dispersion of the pulse.  Ionospheric dispersion causes the pulse to arrive later at lower frequencies, effectively causing the pulse to broaden in time.
The dispersion is proportional to the total
column density of electrons, the Total Electron Content (TEC). TEC is a meteorological phenomenon, and it changes continuously, but most strongly during 
sunrise and sunset. Relevant to the present discussion is the slanted TEC (STEC) which is the TEC value in a slanted column along the observer's line of sight.  STEC and TEC are usually expressed in terms of TEC units (TECU) where $ 1\,\mathrm{TECU} = 10^{16}~ $electrons/m$^2$. The phase shift at a particular frequency is given by
 \begin{equation}
  \phi(\nu) \approx 2\pi \frac{1.34\cdot 10^9 STEC}{\nu} \;.
 \end{equation}
In order to correct for dispersion, we must have a good measure of the STEC which caused the dispersion.  An estimate of the STEC value is available from GPS observations with a 
precision of about $\pm 1$\,TECU. 
For LOFAR, it is likely that the STEC value can be determined even more precisely using images from point sources or Faraday rotation. 
Note that CEP will only use data from the core stations to determine the trigger.  This means that only the STEC at these stations is relevant to triggering. The core stations cover an area of $2\times
2$\,km$^2$, not a large area as far as 
ionospheric phenomena are concerned. This means that local variations of TEC can be ignored.  A single STEC value will be sufficient for de-dispersion of all the core LOFAR stations (see \figref{flowchart}).


\begin{figure}[!ht]
 \centering{
 \includegraphics[width=.32\textwidth]{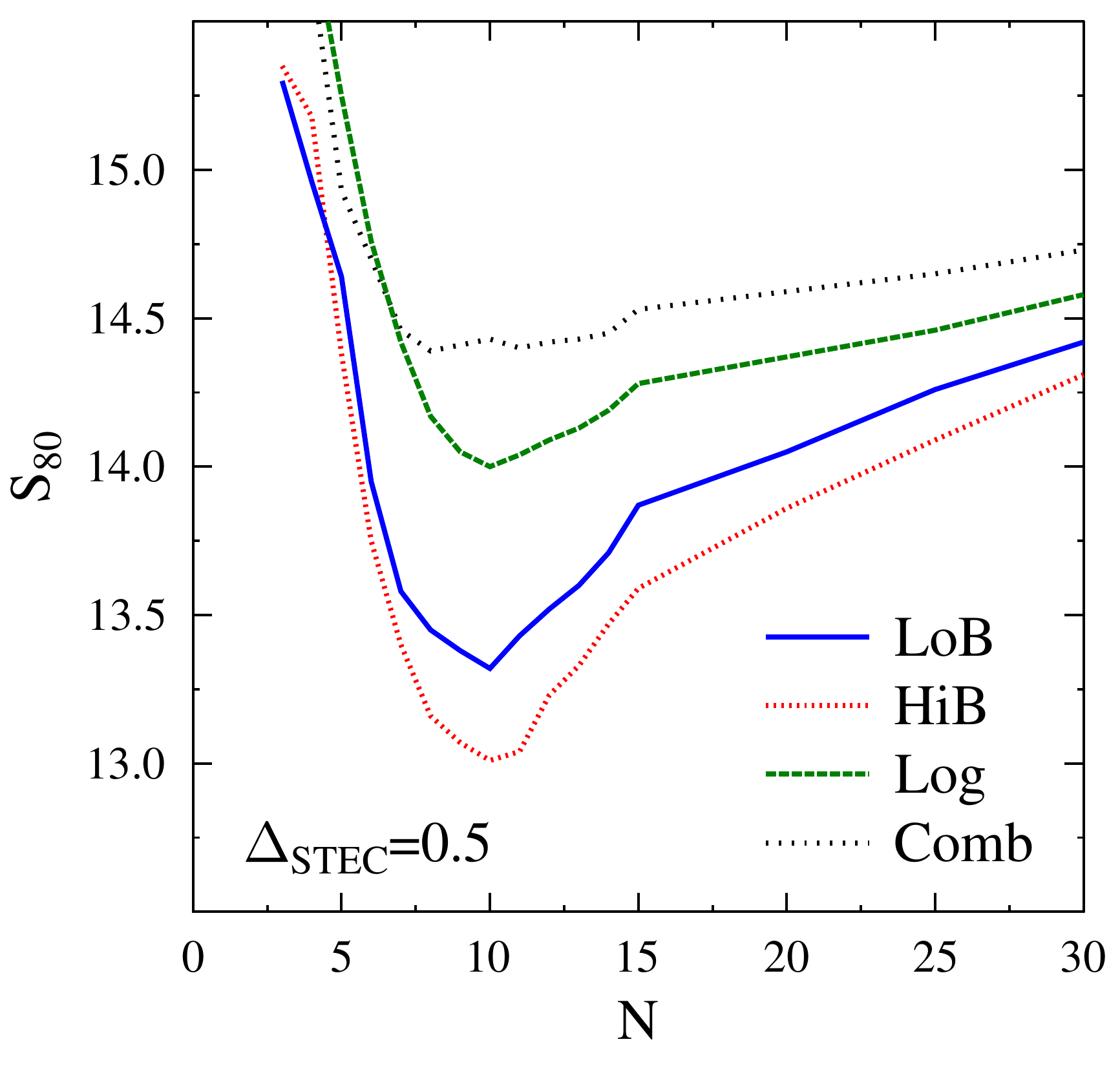} \
 \includegraphics[width=.32\textwidth]{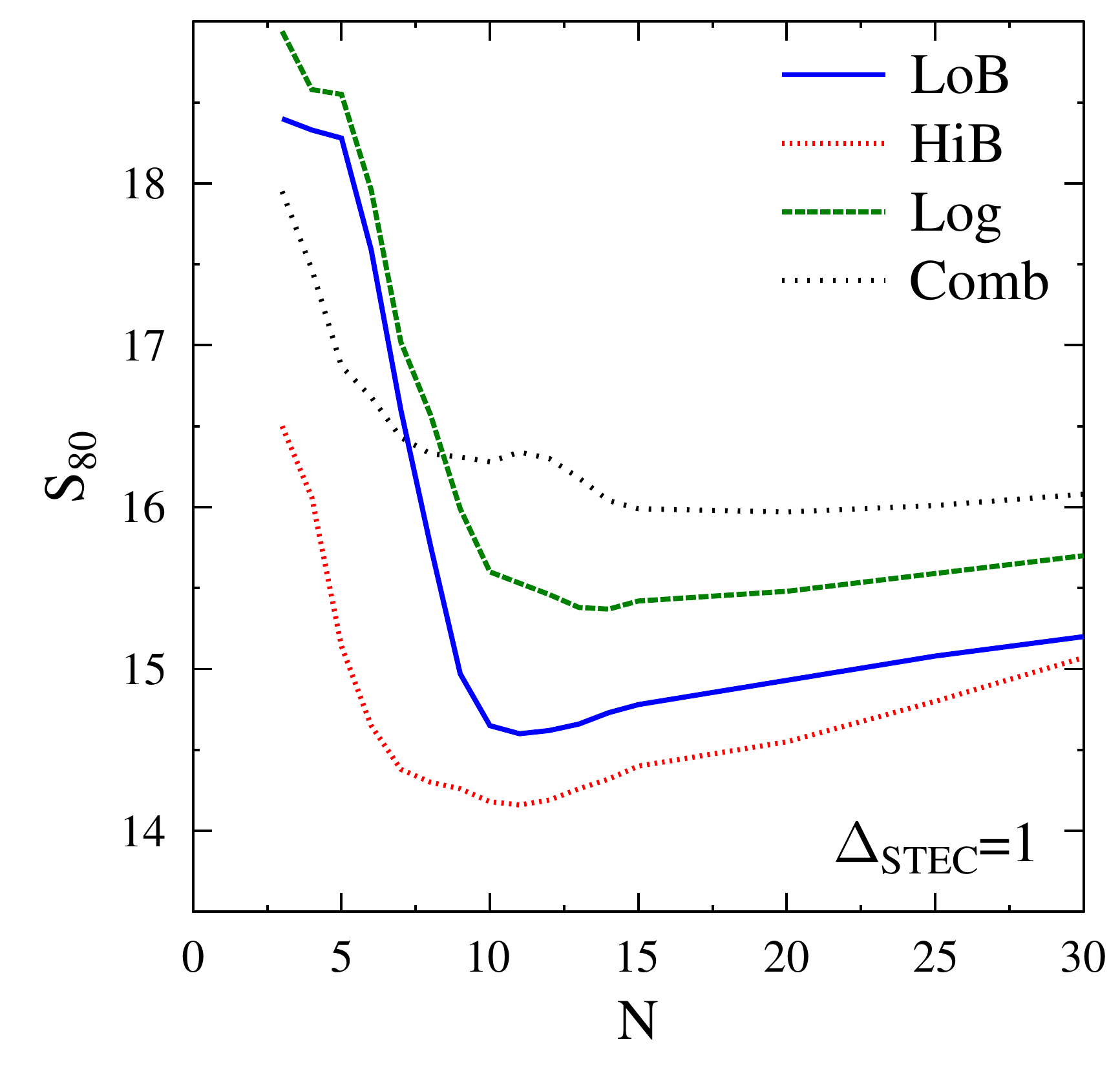} \
 \includegraphics[width=.32\textwidth]{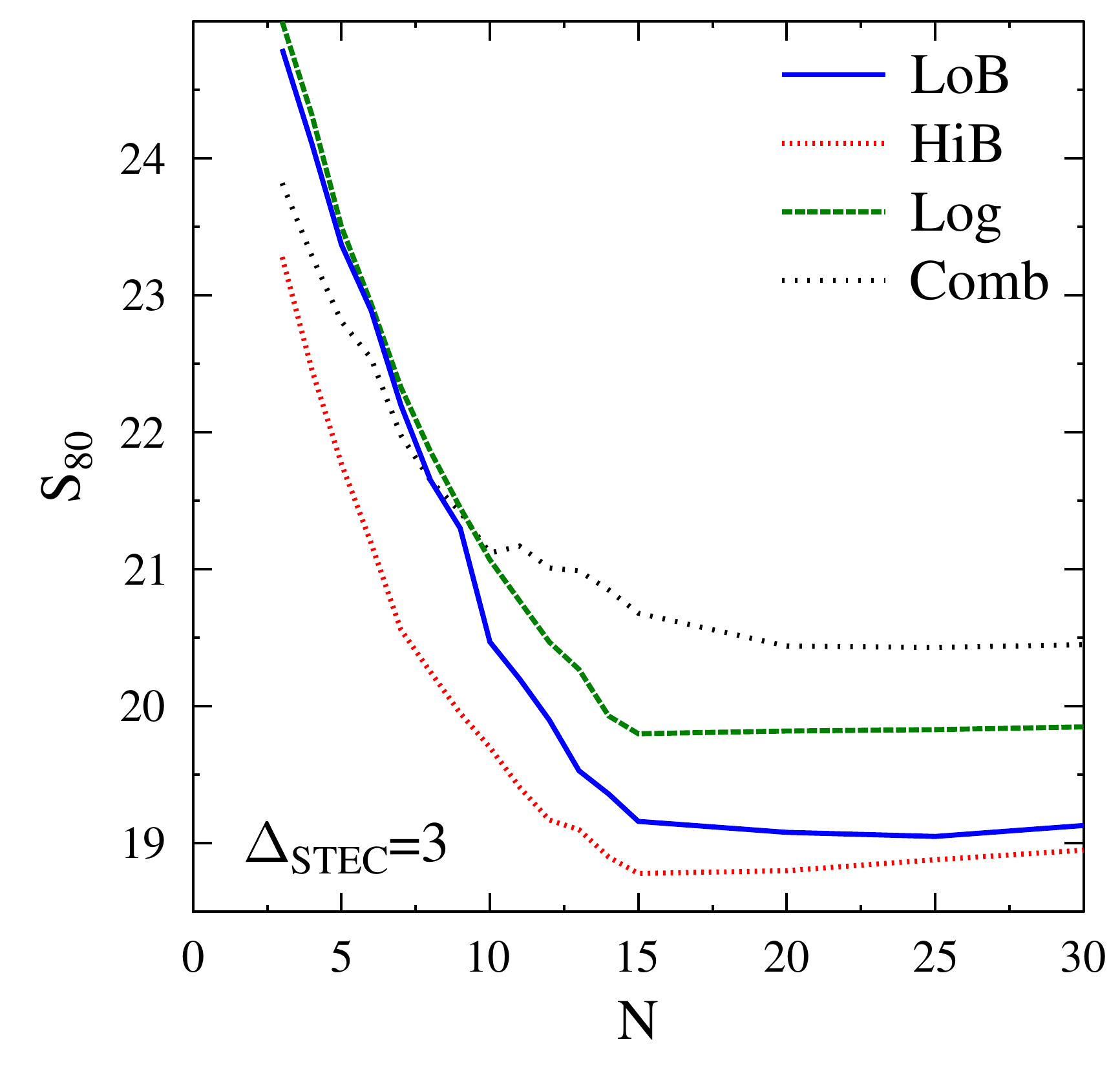}}
 \caption{Same as \figref{de_TimeBins} now including a gaussian spread in the STEC error with a standard deviation increasing from left to right, taking the values of $\Delta_{\rm STEC}=$0.5, 1.0, and 3.0 TECU respectively.}
 \figlab{de_TimeBins_STEC}
\end{figure}

To determine the accuracy to which the STEC needs to be known to perform the proposed Lunar measurements, we have repeated the previous analysis taking a particular STEC value, termed simTEC (=8\,TECU in this case), to disperse the pulse that is added to the data. In the analysis step the pulse is recovered taking different values of the STEC to simulate an error. The difference between the two STEC values is taken according to a Gaussian distribution with width $\Delta_{\rm STEC}$. The results of these calculations are shown in \figref{de_TimeBins_STEC} for different values of $\Delta_{\rm STEC}$. These figures should be compared with the results displayed in \figref{de_TimeBins}. One notices that some clear trends in the plots. The $S_{80}$ values at small values of $N$ rapidly increase. This is can easily be understood from the fact that the uncorrected part of the dispersion of the signal introduces a lengthening of the pulse for an increasing number of time samples reducing the recovered power in a small time window. The same argument also explains that the optimum value for $N$ increases towards larger values with increasing $\Delta_{\rm STEC}$. As a result the optimum value for the window size increases towards larger values at increasing values of the pulse strength that can be recovered with a good efficiency, $S_{80}$.

On the basis of these simulation one thus concludes that for the real observations one should strive to determine the STEC value within an accuracy of $\Delta_{\rm STEC}$=0.5\,TECU. In the simulations to determine the sensitivity for detecting UHE neutrinos we will assume $\Delta_{\rm STEC}=$1.0\,TECU and take $N=15$ to be on the conservative side.

\section{Beaming}\seclab{Beam}

In this section the pulse-detection algorithm is integrated in a realistic antenna configuration where we consider the aspects of beaming. A beam profile is calculated which differs from the usual profiles in the sense that this profile reflects the detection efficiency of broad-band transients. In the calculations the profiles of the tile beams have not been folded in.

\begin{figure*}[!th]
 \centering
 \includegraphics[width=0.7\textwidth]{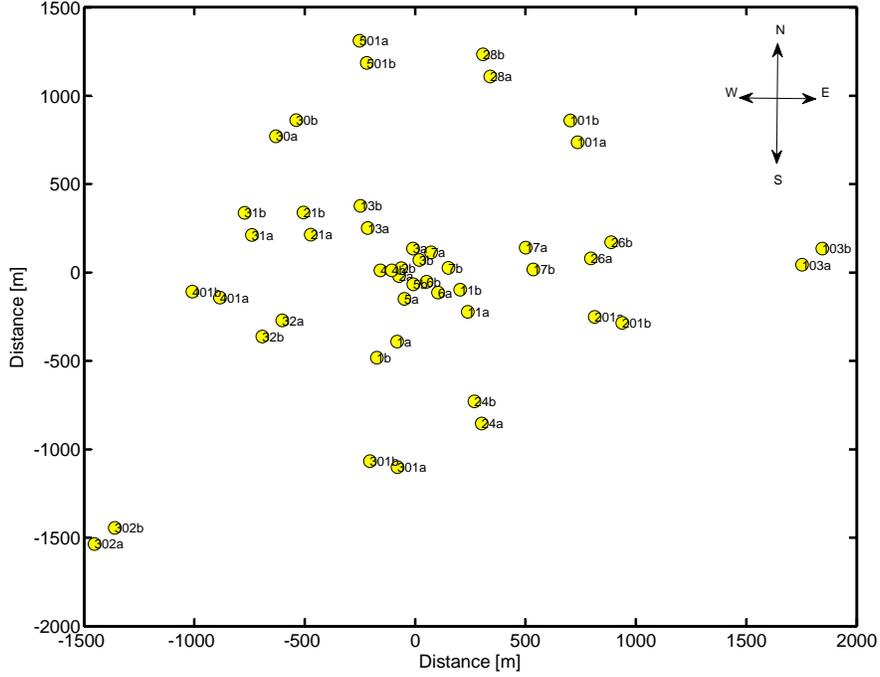} 
 \caption{Layout of the HBA fields of the LOFAR core. Each station consists of two fields.}
 \figlab{corestations}
\end{figure*}

\figref{corestations} shows the layout of the LOFAR core stations. The fields of HBA tiles for every station are shown by yellow circles. Each field represents a group of 24 HBA tiles.
The synthesis of beams using all core stations is required to reach a high sensitivity.

\begin{table}[!ht]
 \caption{Table showing beam widths of LOFAR tied array beam for various position of Moon in the sky, defined by zenith ($\theta$) and azimuth ($\phi$) angles in degrees.
 Simulation are done for the LoB-FFS using all 24 LOFAR core stations. Beam widths are given as $\Delta_\theta$, $\Delta_\phi$. }
 \tablab{lower-frequency-table}
 \begin{center}
 \begin{tabular}{|c|c|c|c|c|c|}
 \hline
 $\phi$ & 120 & 150 & 180 & 210 & 240 \\
 $\theta$ & & & & & \\
 \hline
 15 & 0.07 & 0.072 & 0.076 & 0.078 & 0.07 \\
   & 0.0756 & 0.071 & 0.069 & 0.068 & 0.069 \\
 \hline
 30  & 0.078 & 0.08 & 0.086 & 0.086 & 0.082 \\
   & 0.075 & 0.071 & 0.069 & 0.068 & 0.069 \\
 \hline
  45  & 0.96 & 0.098 & 0.104 & 0.106 & 0.11 \\
    & 0.075 & 0.071 & 0.069 & 0.068 & 0.0693 \\
 \hline
  60  & 0.139 & 0.137 & 0.148 & 0.15 & 0.142 \\
    & 0.077 & 0.071 & 0.069 & 0.067 & 0.069 \\
 \hline
  75  & 0.264 & 0.268 & 0.288 & 0.292 & 0.276 \\
    & 0.075 & 0.072 & 0.069 & 0.068 & 0.069 \\
 \hline
 \end{tabular}
 \end{center}
 \end{table}

Beam widths (FWHM) in the zenith and azimuth angles of tied array beams are indicated for various positions of the Moon in \tabref{lower-frequency-table} for the LoB-FFS.
The azimuth angle variation is taken from 120$^\circ$ to 240$^\circ$ (where $\phi=0^\circ$ is north and $\phi=90^\circ$ is west) to match the moonrise and moonset directions.
The values given in the table can easily be understood from the fact that at $\phi \approx 120^\circ$ the station layout \figref{corestations} gives the smallest projected baseline while the largest baseline is seen at $\phi \approx 210^\circ$. The azimuth beamwidth $\Delta\,\phi$ should thus be largest at $\phi = 120^\circ$ and smallest at $\phi = 210^\circ$ as shown by the numbers in the table. Simple geometry shows that $\Delta_\phi$ should be independent of zenith angle.
Furthermore from geometry one deduces that the beamwidth in zenith angle at ($\theta$, $\phi$) equals $\Delta\,\phi$($\theta$, $\phi-90^\circ$)/$\cos{\theta}$.

We will cover the whole Lunar surface with several beams that overlap at the FWHM angle.
From the beamwidth given in \tabref{lower-frequency-table} it can be calculated that thus 48 beams are necessary to cover the whole visible Lunar surface (an angular size of half a degree) if it were at Zenith (which it never is). At finite zenith angle $\theta$ the angular area of the coherent beams increases and fewer beams, 48$\times \cos{\theta}$, independent of azimuth angle, are necessary to cover the area of the Moon.

The observed beam widths are frequency dependent. For the HiB-FFS, where the wavelengths are shorter, the widths are more narrow and one finds $\Delta\,\phi=0.062^\circ$ and $0.056^\circ$ as maximum and minimum at $\phi = 120^\circ$ and $210^\circ$ respectively.

\begin{figure}[!ht]
 \centering
 \includegraphics[width=.45\textwidth]{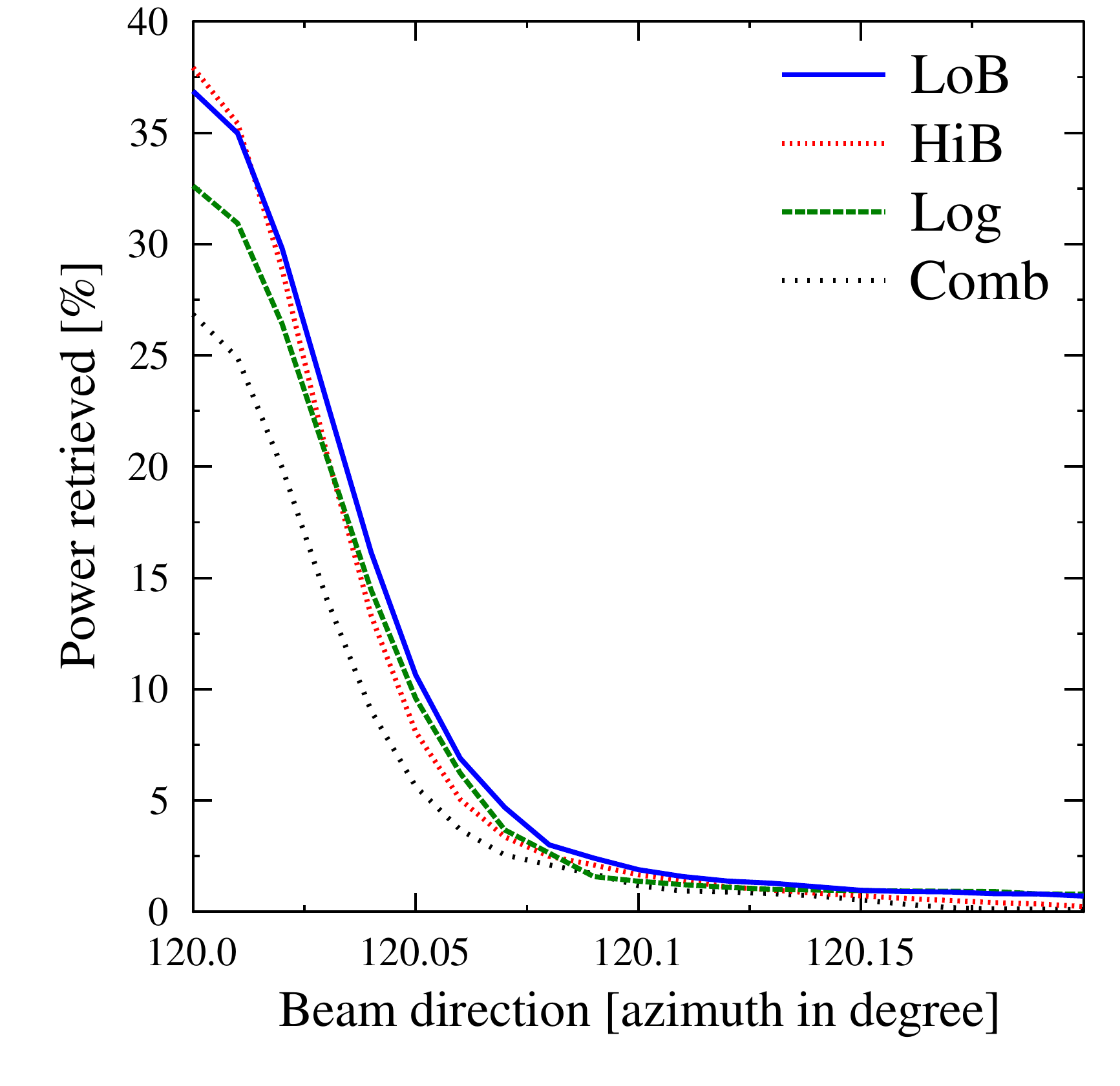}  
 \caption{Detection efficiency along azimuth angle $\phi$ for pulses de-dispersed with no STEC error
 when the source of the pulse is assumed to be at ($\theta=60^\circ$, $\phi=120^\circ$).}
 \figlab{azimuth_beamfilt}
\end{figure}

\begin{figure}[!ht]
 \centering
 \includegraphics[width=.45\textwidth]{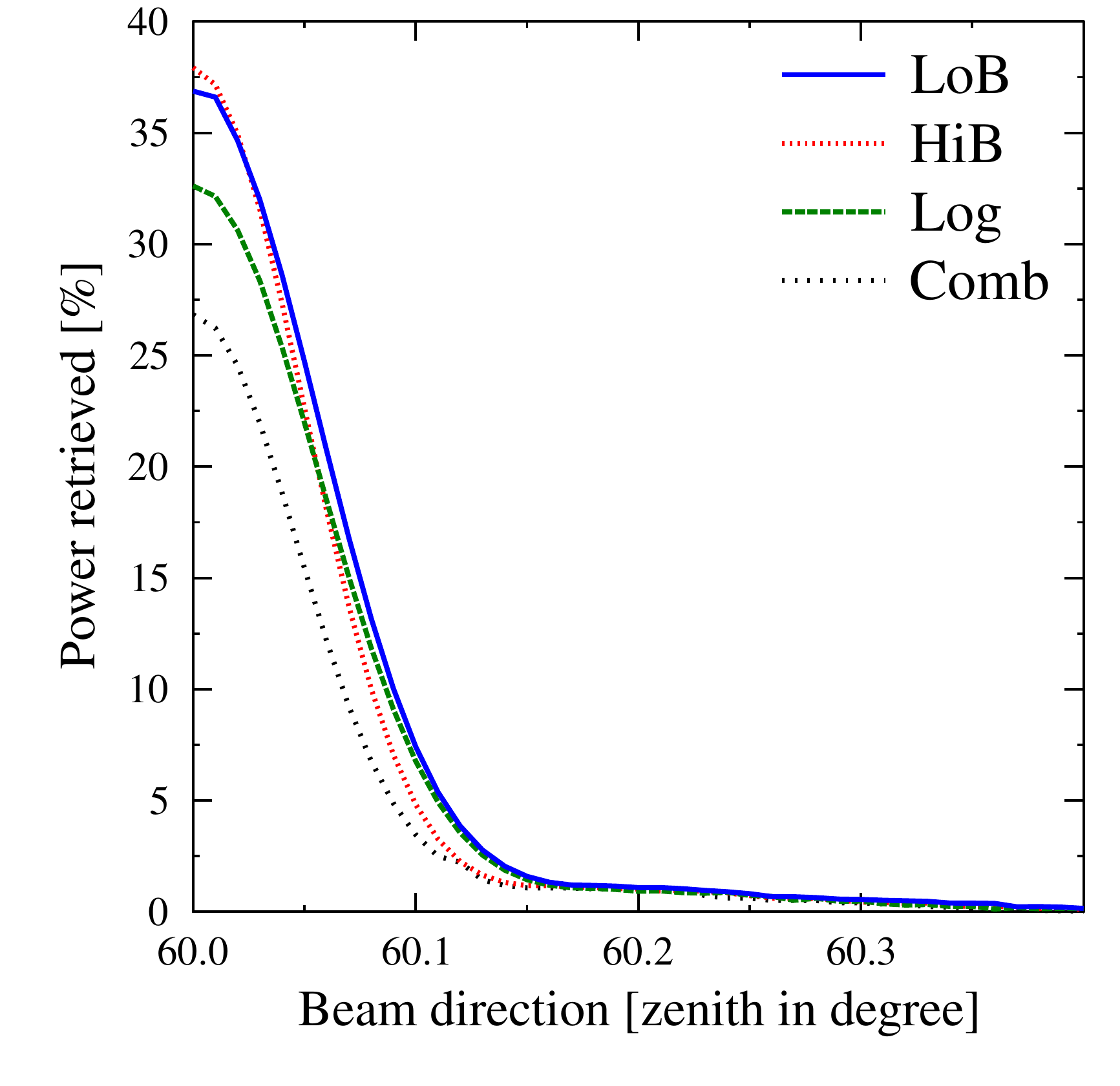}  
 \caption{Same as \figref{azimuth_beamfilt}, but for the detection efficiency along zenith angle $\theta$.}
 \figlab{zenith_beamfilt}
\end{figure}

An important ingredient of the trigger software is the implementation of an anti-coincidence requirement that will suppress a large fraction of the transient noise. For this it is necessary to investigate the magnitude of the side-lobes for the pulse-response.
We have run simulations to model the response to different source directions for the LOFAR core configuration. A pulse is added to the time traces of the different core stations as arriving from a certain direction, ($\theta=60^\circ$, $\phi=120^\circ$). In the reconstruction the signals are added with phases corresponding to a slightly different viewing direction. We have not included noise in this simulation as it is not essential. The full trigger pipeline was simulated. The results for $P_N^\mathrm{m}$ using $N=15$ for the different FFSs are shown in \figref{azimuth_beamfilt} and \figref{zenith_beamfilt}. From these figures it is seen that the sidelobes are strongly suppressed due to the (almost) random relative positions of the core stations.

\section{Energy limit of ultra high energy particle}\seclab{limits}


\begin{figure}[!ht]
 \centering
 \includegraphics[width=.5\textwidth]{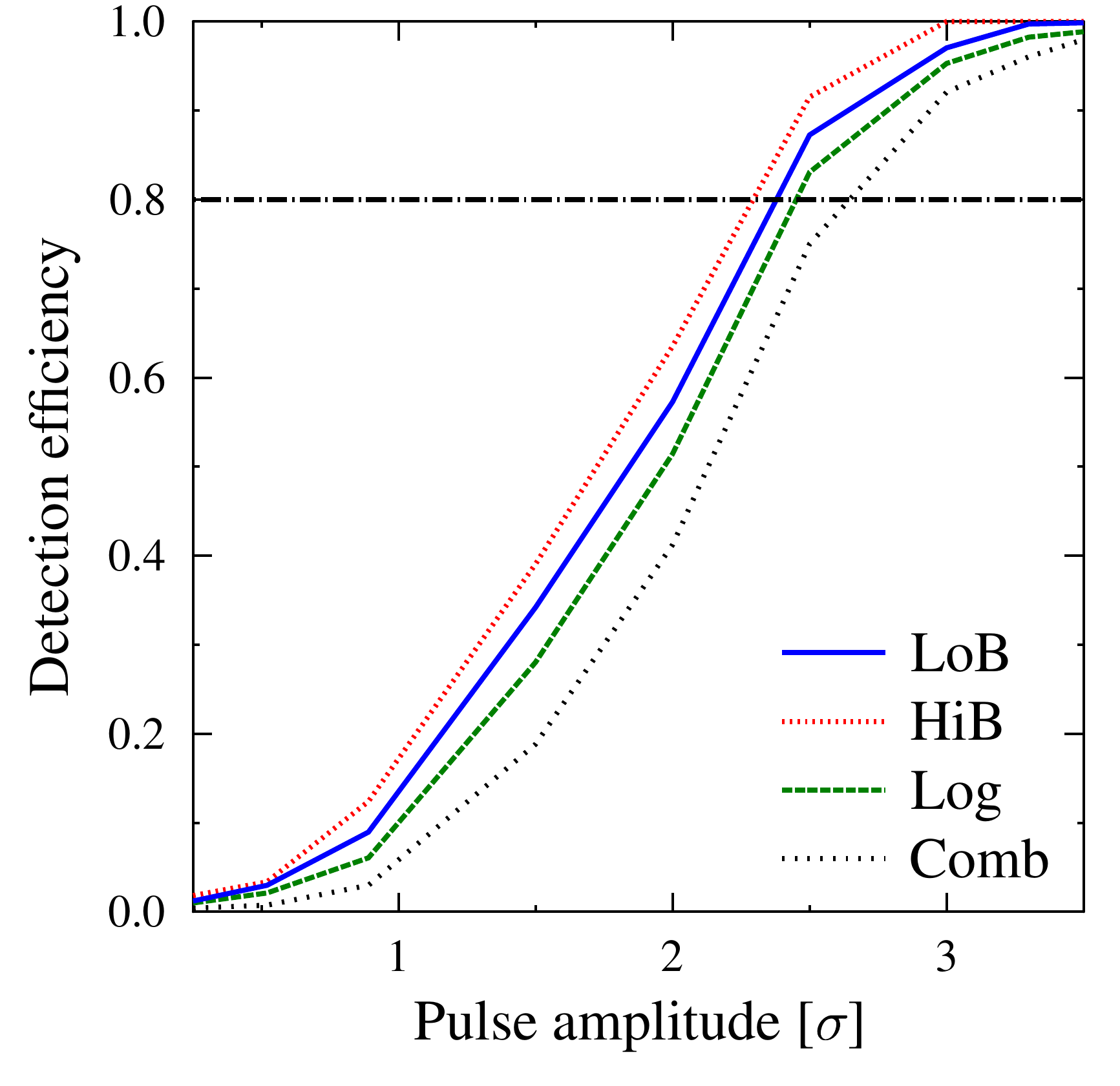}
 \caption{Shown is the detection efficiency for the optimum setting, $N=15$, using the LOFAR 
core configuration and averaging over the full-width at half maximum of the beam. }
 \figlab{de-final}
\end{figure}

The detection efficiency is investigated for the various filter schemes for the  LOFAR core. Pulses are added to a un-correlated Gaussian background and are dispersed using Gaussian distributed values around the mean STEC value 
that is corrected for in the analysis, with a standard deviation of 1 TECU.  The simulations are done for $1000$ added pulses.
The detection efficiency for the optimal settings for the window length,  $N=15$, is shown in \figref{de-final}. This calculation includes the effects of coherent addition of the 24 stations in the core that are already deployed where the source is spread over an angular range corresponding to the size of the beam.


The limit for the trigger rate we want to consider is about once every minute.
This gives, using \figref{de-final}, \tabref{SEFD}, and including a factor $\sqrt{2}$ to account for a linearly polarized signal, an 80\% sensitivity for pulses with an intensity in excess of 26\,kJy. 

In calculating the sensitivity of the LOFAR measurements to pulses coming from the Moon one should realize that the final sensitivity reached in an off-line processing of the data cannot be larger than the trigger level that has been set. Any pulses with lower strength will not set the trigger flag and are thus lost for later processing. The limit considered for this work will be based on a single pulse for the duration of the observations, a few days. In the actual observations one may consider the number of excess counts over a statistical noise distribution, however, this requires a perfect understanding of the transient noise which is the subject of a future work.
The highest sensitivity is reached when the post processing is performed using the full bandwidth information stored on the TBBs while the trigger signal is obtained using the LoB-FFS.

The sensitivity that can be reached in post processing is determined by the accidental rate for the full LOFAR. A safe limit can be set if the accidental rate vanishes for the duration of the complete observation. Setting this ---relatively arbitrarily--- to one month we arrive at a threshold for accidental detection which is increased by a factor of less than 1.5 (using \eqref{Ptk}) as compared to that for one accidental detection per minute (the trigger threshold). The full LOFAR will have a collecting area that is double that of the core, and we will be able to use the full bandwidth giving a factor 4 increase in the signal over noise ratio. The 80\% sensitivity level for pulses thus lies at a much lower value than the trigger value of 26\,kJy.

\begin{figure}[!ht]
 \centering
 \includegraphics[width=.5\textwidth]{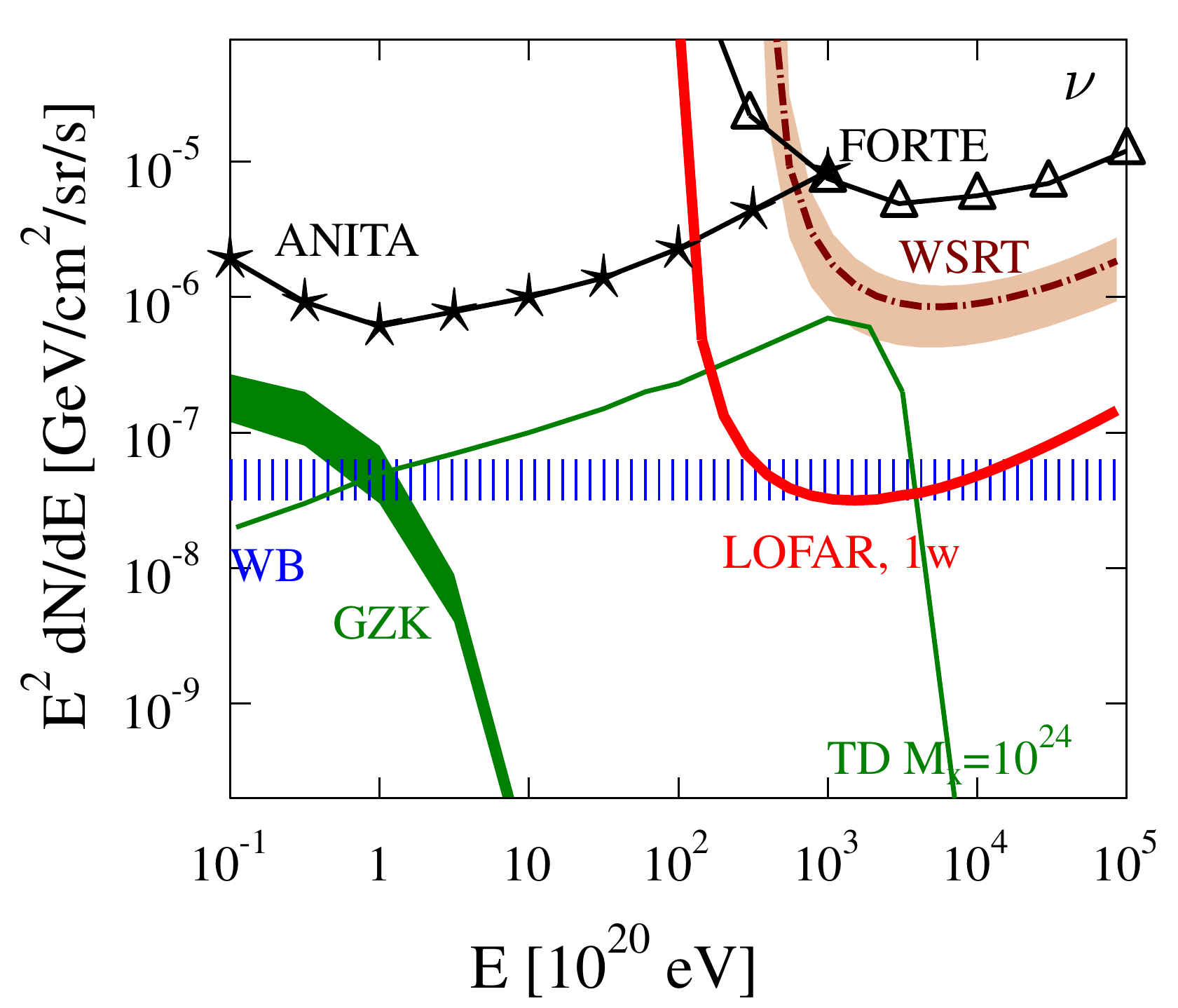}
 \caption{Neutrino flux limits for LOFAR, see discussion in the text. }
 \figlab{LOFAR-nu}
\end{figure}

From these considerations it is clear that the determining factor for the observations is set by the trigger threshold. In \figref{LOFAR-nu} the sensitivity on the neutrino flux for LOFAR is given, based on the pulse-detection thresholds indicated above.
The obtained limit is getting sensitive to the Waxman-Bahcall flux prediction~\cite{Wax98} based on a polynomial extrapolation of the measured cosmic-ray flux and of the order of 40 counts are expected if the predictions of a top-down model~\cite{ps96} for exotic particles of mass $M_X=10^{24}$~eV would be correct.
The previous limits in the UHE region have been set by the ANITA~\cite{anita10} and FORTE~\cite{forte} experiments. The GZK neutrino flux indicated in the figure is obtained from the work of ref.~\cite{GZK-2}.

The detection threshold for the LOFAR observations is more than an order of magnitude lower than the 240\,kJy for the observations with the WSRT~\cite{Bui10}. Since the strength of the pulse generated by the neutrinos depends quadratically on the energy, the LOFAR observations are sensitive to neutrinos with much smaller energies, as can be seen from \figref{LOFAR-nu}. The increased sensitivity, combined with the longer observation time makes the observations sensitive to considerably lower neutrino fluxes.

\begin{figure}
\centering
\includegraphics[width=.3\linewidth]{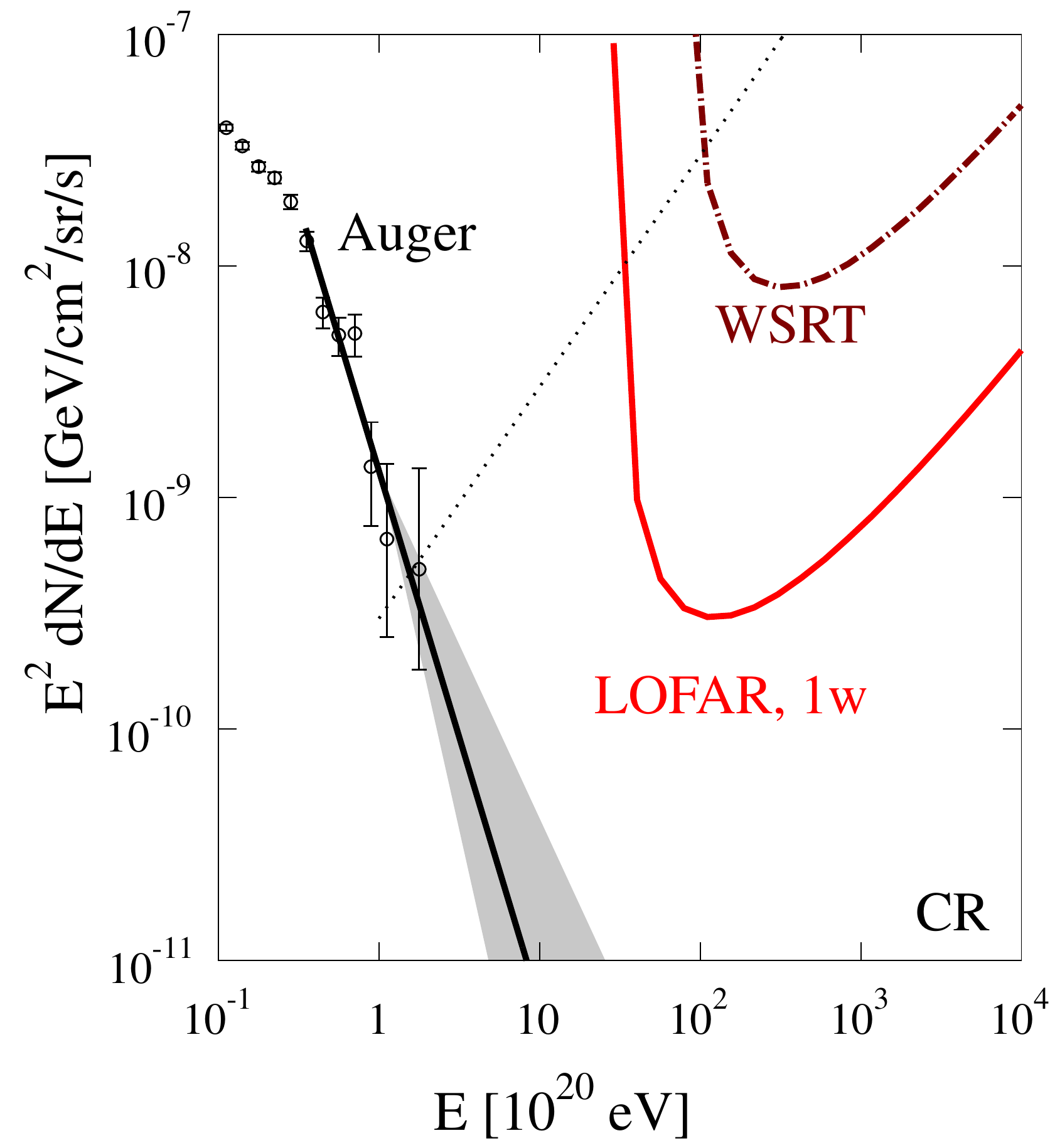}
 \hspace{1em}
\begin{minipage}[b]{.65\textwidth}
\caption{ The limit for the cosmic-ray flux as can be determined by LOFAR in one week observation time is compared to the flux determined by the Pierre Auger Observatory~\cite{Abraham:2010mj} (data points with error bars) and a simple polynomial expansion (black line, see text). Also the prospective flux sensitivities are indicated that can be obtained with LOFAR. Shown is also the cosmic-ray flux limit from WSRT observations~\cite{Bui10,terV10}.}
\figlab{LOFAR-cr}
\end{minipage}
\end{figure}

Short radio pulses emitted from the lunar regolith can also be used to detect UHE cosmic rays. The main differences between the interactions of cosmic-rays and neutrinos in the regolith is that in cosmic rays all the energy is converted into a particle shower while this is only of the order of 20\% for neutrinos. Another important difference is that while neutrino showers develop deep inside the regolith, cosmic ray showers develop very close to the Lunar surface. Recently is has been shown that showers close to vacuum-medium boundary emit electromagnetic radiation to the same extent as would be obtained by using plane-wave refraction of the waves waves through the surface~\cite{terV10}. As shown in \figref{LOFAR-cr} this allows for a tightening of the flux limits at the highest energies, well below the model-independent limits extracted from the data obtained at the Pierre Auger Observatory~\cite{Abraham:2010mj}.

\section{Summary and conclusions}

As an essential part of the project to determine the flux of UHE particles through their impacts on the Lunar surface, we have investigated the most efficient method to detect the radio pulse, emitted by the impact, in a noisy background. Since the data processing is performed in real time the calculational latency of the method must be small. To be able to handle the enormous data rate generated by LOFAR we propose a procedure that consists of two separate stages. The first stage generates a trigger signal based on limited information available from the core stations. The trigger causes the complete, full bandwidth, signal from the core as well as the remote stations to be written to a mass storage system for later processing. In a second processing stage the stored data will be searched for Moon pulses. At this stage full bandwidth and the maximum 
collecting area 
are available for analysis and thus the ultimate sensitivity can be reached for pulse detection. This leaves the construction of the trigger signal as the defining bottleneck in the system.

To limit the latency in constructing the trigger we have restricted ourselves to procedures where the power in a time window is compared to a threshold value. Of particular importance in determining the window size is the selection of the frequencies used in the construction of the trigger, as only half the bandwidth can be processed. Another important consideration is the accuracy with which the signal can be corrected for the dispersion caused by the free electrons in the upper ionosphere. These factors are taken into account in a simulation and optimal trigger conditions are determined. On the basis of these optimal settings the sensitivity of observations to Lunar pulses is determined which translate into flux limits. This shows that observations with LOFAR, in the frequency range of 100--200\,MHz are an order of magnitude more sensitive than previous observations in this frequency range looking for Lunar pulses.

\section{Acknowledgements}

This work was performed as part of the research programs of the Stichting voor Fundamenteel Onderzoek der Materie (FOM), with financial support from the Nederlandse Organisatie voor Wetenschappelijk Onderzoek (NWO), and an advanced grant (Falcke) of the European Research Council.
\Omit{"LOFAR, the Low Frequency Array designed and constructed by ASTRON, has facilities in several countries, that are owned by various parties (each with their own funding sources), and that are collectively operated by the International LOFAR Telescope (ILT) foundation under a joint scientific policy." }
 \appendix

\section{The polyphase filter and inversion}\seclab{PPFInv}

The fast fourier transform is a very efficient method for transforming the data between the frequency and time domains. However, the resolution in frequency is limited.  Without the application of a windowing function (such as a Hamming filter), this causes considerable leakage of signal between neighboring channels in the frequency domain.  On the other hand, if a windowing function is applied, considerable intensity is lost.
One way to overcome this problem is to work with overlapping the sections of the data incorporated as a polyphase filter bank.  The PPF banks are implemented on FPGAs (Field Programmable Gate Arrays) at the LOFAR stations (see \figref{flowchart}).

The PPF bank is a combination of a parallel structure of $M$ ($M=1024$ for the present implementation) sub-filters followed by an FFT stage~\cite{Mitra}.
Each sub-filter is a Finite Impulse Response (FIR) filter (like the main filter) that filters with $K=16$ taps (or filter coefficients). The total filter-structure can be represented as a matrix with $M$ rows and $K$ columns where each sub-filter is fed with input data $M$ time samples apart.
The weighted average of $K$ input time samples will be summed and fed into the $M$ point FFT.

\begin{figure}[!ht]
 \centering
 \includegraphics[width=.5\textwidth]{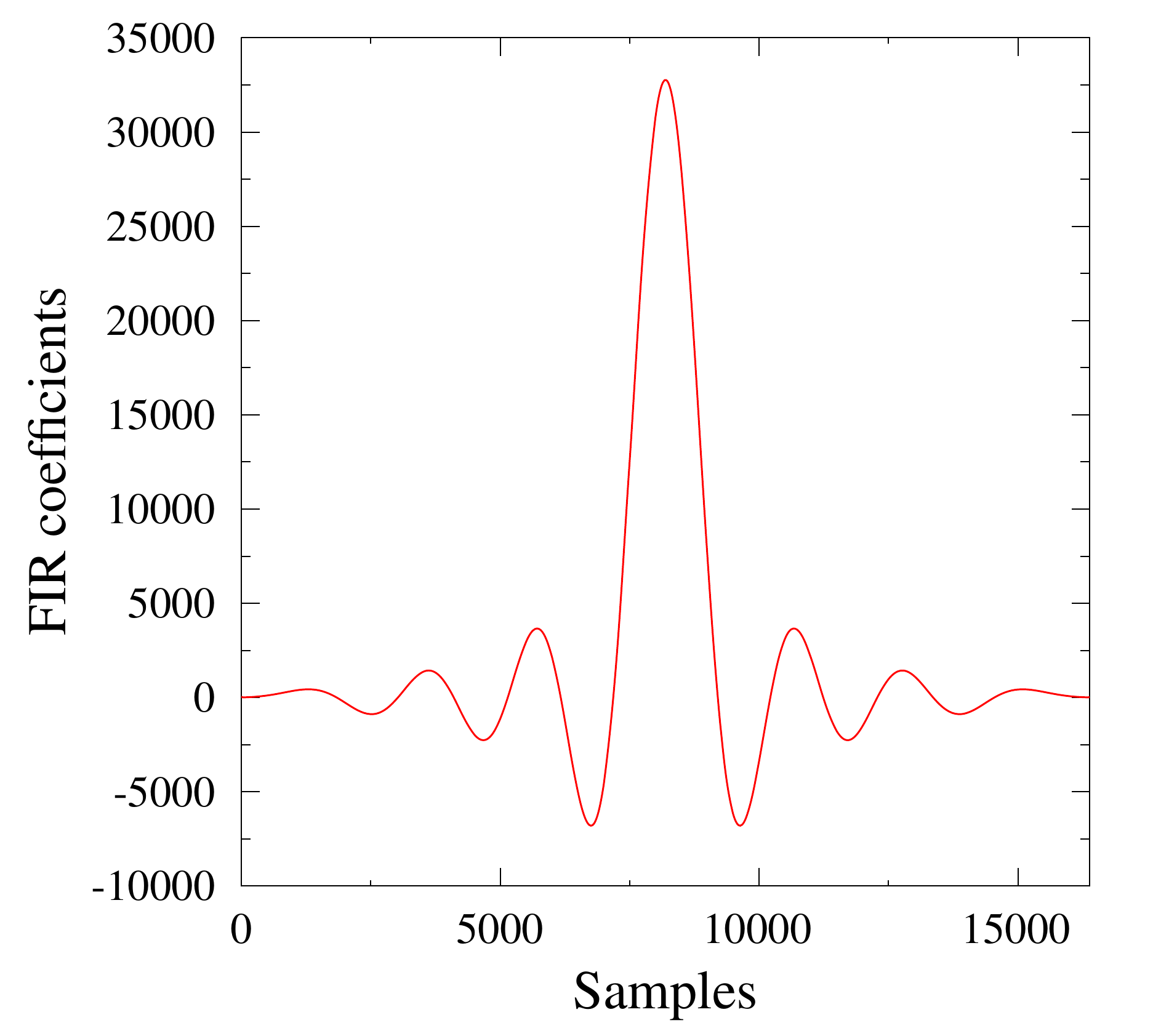}
 \caption{Impulse response of the implemented FIR filter.}
 \figlab{transfer-function-ppfcoeff}
\end{figure}

The impulse-response of LOFAR's FIR filters is similar to a sinc-function, which inherits the linear-phase of each  subband~\cite{Mitra}. The implemented impulse-response with all $M \times K$ sub-filters is shown in \figref{transfer-function-ppfcoeff}.

\begin{figure}[!ht]
  \centering
  \subfloat[PPF Spectrum]{\figlab{PPFspectrum}\includegraphics[
  width=0.49\textwidth]{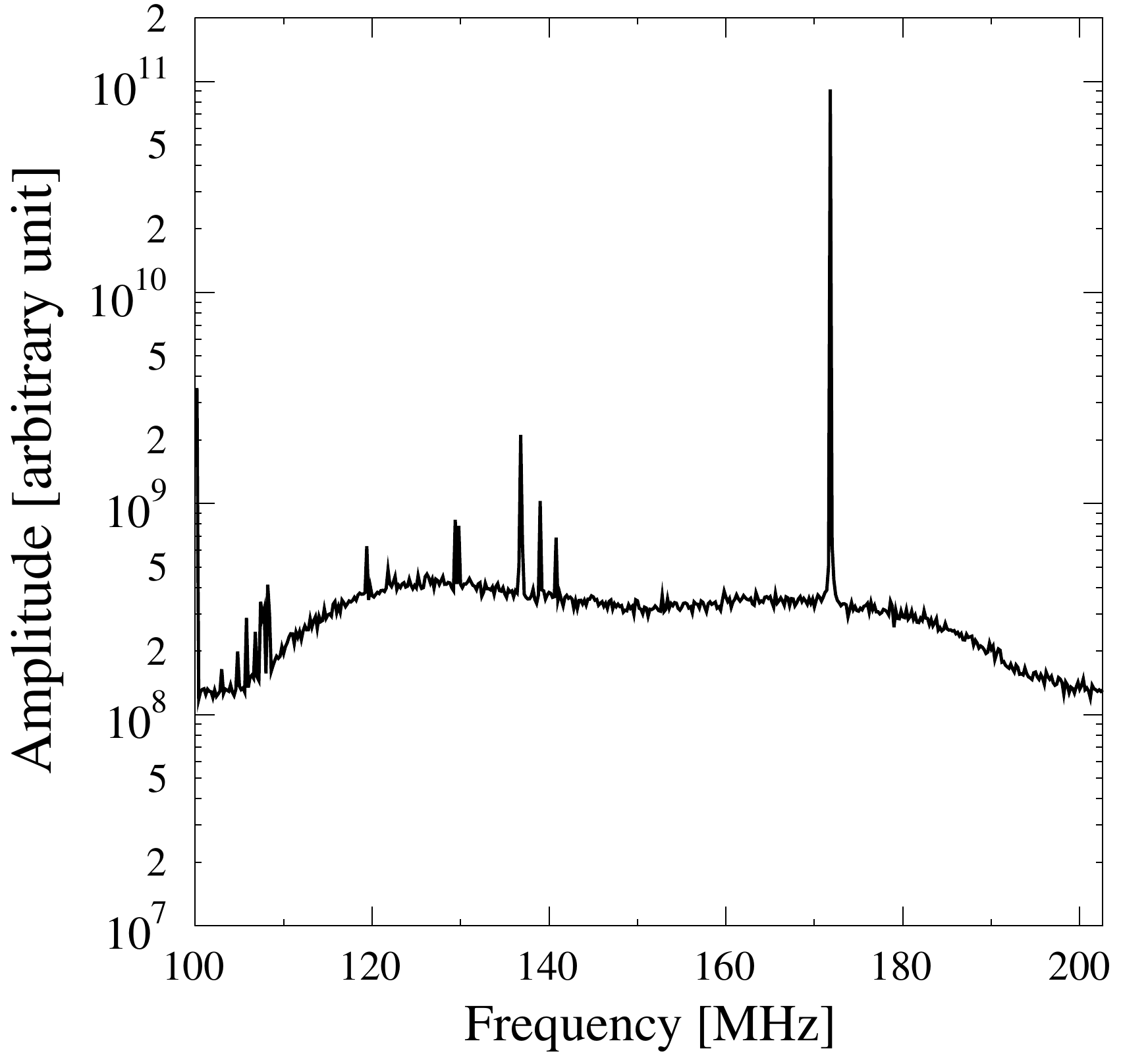}}
  \subfloat[FFT Spectrum]{\figlab{FFTspectrum}\includegraphics[
  width=0.49\textwidth]{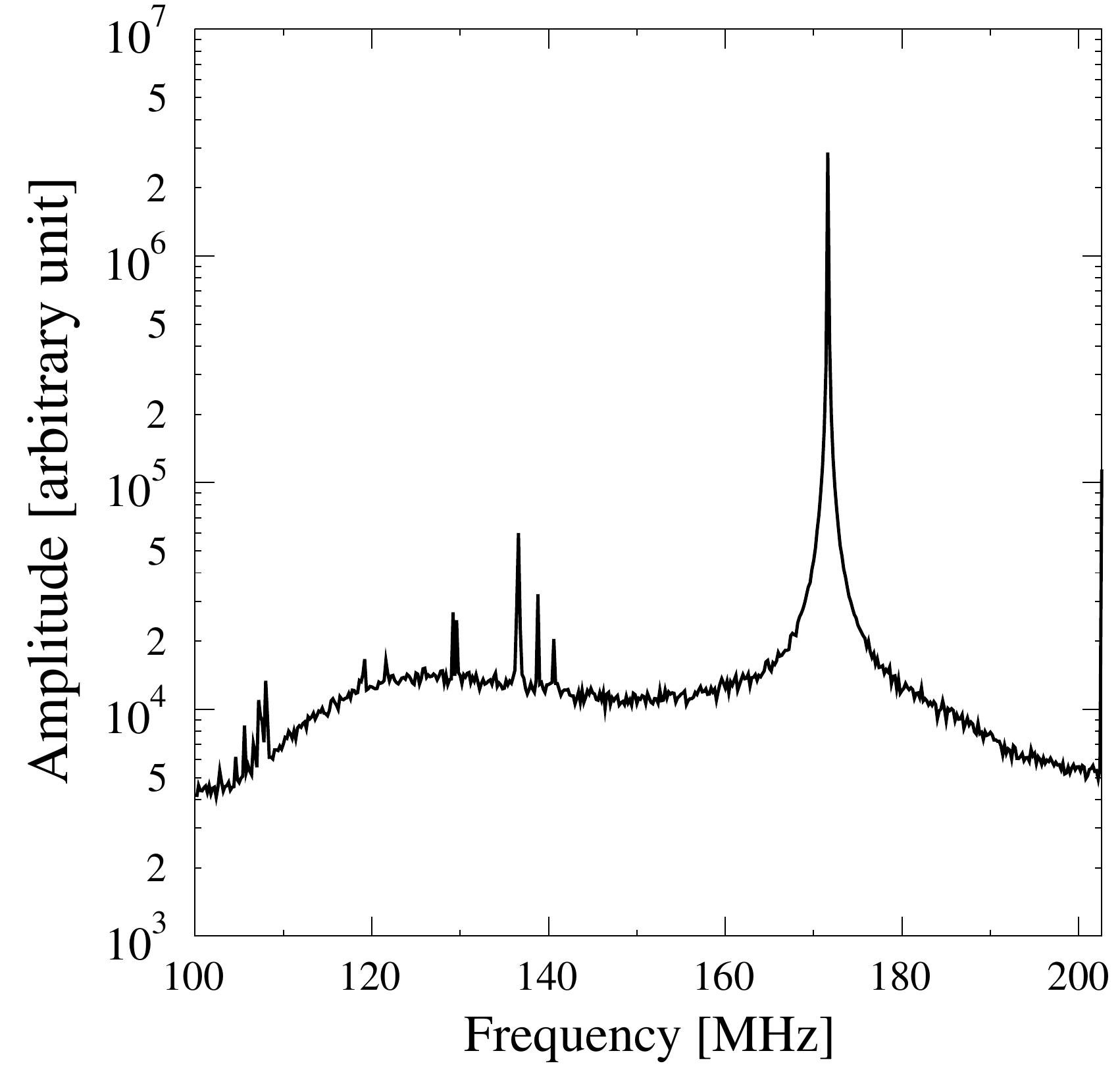}}
  \caption{The frequency spectrum  of an HBA tile of LOFAR as determined from the PPF bank is compared with that of a simple FFT.} \figlab{PPFFTspectrum}
\end{figure}

The advantage of using the PPF can be seen from the frequency spectrum of the HBAs of LOFAR (\figref{PPFFTspectrum}) where there is a strong NRFI at 169.65 MHz. The spectrum on the r.h.s.\  of \figref{PPFFTspectrum} is obtained by performing an FFT transform on a block of 1024 time samples.
This shows that using a PPF is a very efficient way to suppress aliasing of NRFI lines to adjacent subbands which is important for efficient NRFI mitigation, as discussed in \secref{RFIMitigation}.

Because the data stream is split into frequency subbands by the PPF, efficient online  beamforming and STEC-correction is possible.
For triggering, we must reconstruct the original time-domain signal by performing an inversion of the action of the PPF. The PPF inversion routine (PPF$^{-1}$) is implemented on CEP (see \figref{flowchart}). Since exact inversion leads to instabilities the inversion algorithm is based on a Least Mean Squares (LMS) Filter approximation for the inversion. A LMS Filter is an adaptive filter that adjusts its transfer function according to an optimized algorithm. The method for FIR inversion is as follows.  The filter is provided with an example of the desired output, together with the corresponding input signal.  The filter then calculates the filter weights (coefficients) that produce the least mean squares fit to the input signal.  In this case, we have calculated the time-domain inversion of an impulse response (transfer function) for all $M$ sub-filters of the PPF bank.

\begin{figure}[!ht]
 \centering
 \includegraphics[width=.5\textwidth]{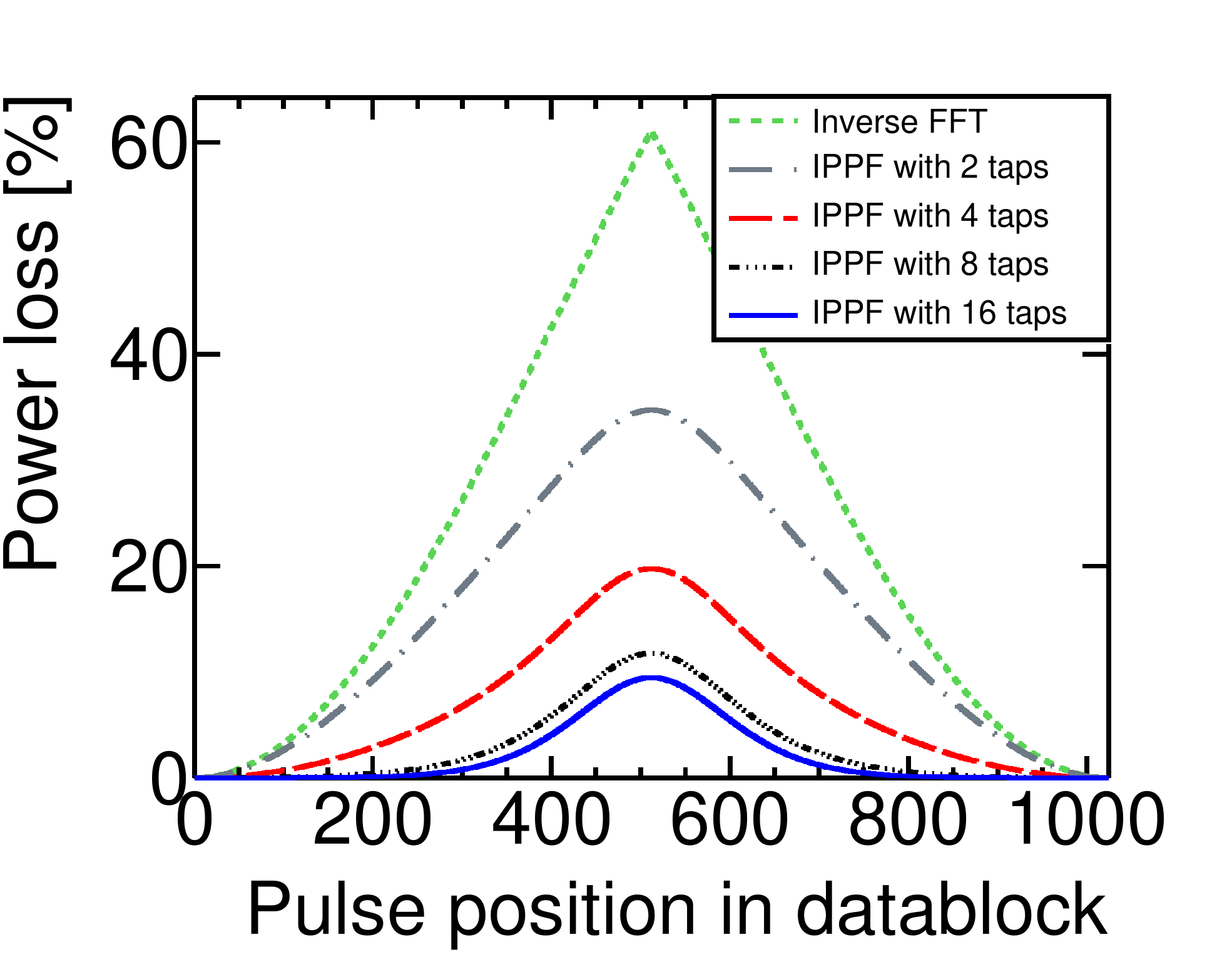}
 \caption{The power loss for pulse reconstruction as a function of position of short pulse in a
 data block of 1024 samples. The pulse is reconstructed using a simple inverse FFT (dashed, green), an PPF$^{-1}$ with 2 taps (dotted, green), an PPF$^{-1}$ with 4 taps (long dashed, red), and an PPF$^{-1}$ with 8 taps (dashed-dotted, black)  instead of PPF$^{-1}$ of 16 taps (drawn, blue).}
 \figlab{pulserecovery}
\end{figure}

It is computationally expensive to implement the PPF inversion, because it increases the latency in the online data processing.
We have considered using a fewer number of taps in the PPF inversion in order to to save CPU-processing time.
To test this a Nyquist-sampled pulse was placed at an arbitrary position in a page of $1024$ time samples.  The PPF transformation (with 16 taps) and its inversion (with a smaller number of taps) was implemented on the simulated pulse.
\figref{pulserecovery} shows the percentage of power loss in the reconstructed pulse as a function of position of the simulated pulse in the page. The power of the recovered pulse is obtained by integrating over $20$ time samples.
The length of the PPF equals an even number of pages. Since the signal reconstruction is optimal for a pulse in the center this explains why the efficiency shown in \figref{pulserecovery} is best near the edges of the page.
For the full PPF inversion with 16 taps the power loss is approximately $10\%$ when the pulse is in the center of the page.  The loss strongly increases when the pulse is recovered using a smaller number of taps.   For these same pulse-positions, loss approaches $20\%$ for $4$ taps, and is nearly $35\%$ for $2$ taps.  Note that a simple inverse FFT is equivalent to a PPF inversion that is done without applying the inverse FIR filter function. In this case the power loss reaches 60\%. We thus conclude that reducing the number of taps in the PPF inversion routine results in a considerable loss of intensity for the pulse-response of the system.

\begin{figure}[!ht]
 \centering
 \includegraphics[width=.5\textwidth]{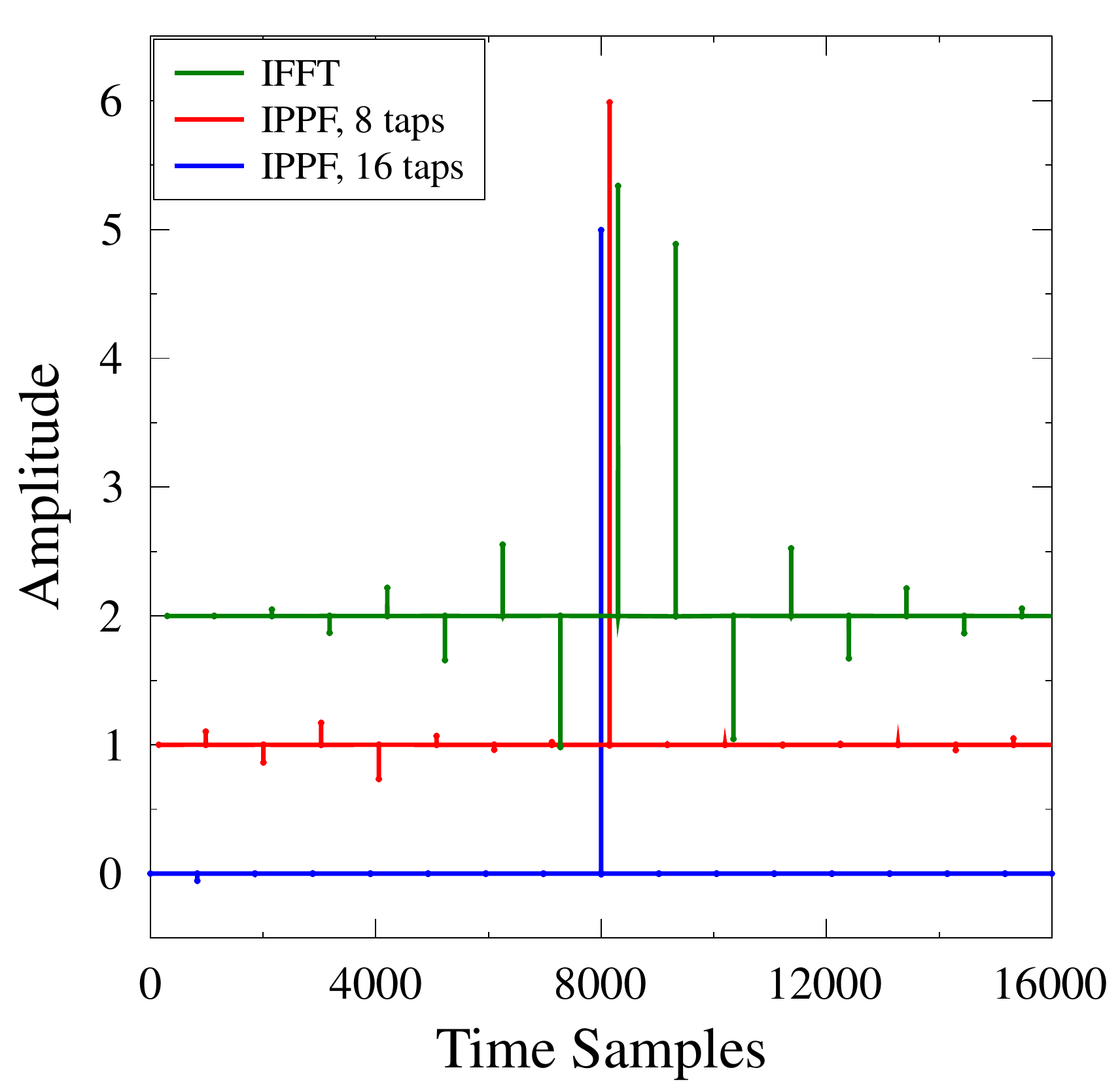}
 \caption{Inverted delta pulse structure with a simple inverse FFT (green, offset=2), and PPF$^{-1}$ with 8 taps (red, offset=1) instead of PPF$^{-1}$ of 16 taps (blue). The test pulse has amplitude 5. The spectra are also a little offset in time to increase visibility.}
 \figlab{pulse_info}
\end{figure}

Because of the initial 16-taps, the PPF strength of a pulse is distributed over 16 output signals. Using the inverse PPF with 16 taps re-combines this information to reproduce the original pulse. If the inversion is performed with fewer taps, or (in the extreme) by performing an inverse FFT, the strength of the original pulse is distributed across multiple echos (see \figref{pulse_info}) which get worse when fewer taps are included in the inversion. It should be noted that the Gaussian noise level stays at the same strength when processed this way.  This is because the redistribution of Gaussian noise signals results in both constructive and destructive interference of these noise signals.  By contrast, a single sharp, well-defined pulse cannot experience constructive interference with itself, and will only be reduced by reducing the number of taps.

\end{document}